\documentclass[12pt,preprint]{aastex}
\newcommand{\mpe}{\dot{M}_{pe}}
\newcommand{\macc}{\dot{M}_{acc}}
\newcommand{\mpy}{{\rm M}_{\odot} {\rm yr}^{-1}}
\begin{document}
\title{Photoevaporation of Circumstellar Disks by FUV, EUV and X-ray Radiation from the
Central Star} 
\author{U.~Gorti\altaffilmark{1,2} }
\author{D.~Hollenbach\altaffilmark{2}} 
\altaffiltext{1}{University of California, Berkeley, CA}
\altaffiltext{2}{NASA Ames Research Center, Moffett Field, CA}   
\begin{abstract}
We  calculate the rate of photoevaporation of a circumstellar disk by energetic radiation (FUV, 6eV $<h\nu<$13.6eV; EUV, 13.6eV $<h\nu<$0.1keV; and Xrays, $h\nu>0.1$keV)  from its central star. We focus on the effects of FUV and X-ray photons since EUV photoevaporation has been treated previously,  and consider central star masses in the range $0.3-7 {\rm M}_{\odot}$.    Contrary to the EUV photoevaporation scenario, which  creates a gap at about $r_g\sim 7\  (M_*/1{\rm M}_{\odot})$ AU and then erodes the outer disk from inside out, we find that FUV photoevaporation predominantly removes less bound gas from the outer disk.  Heating by FUV photons can cause significant erosion of the outer disk where most of the mass is typically located.  X-rays indirectly increase the mass loss rates (by a factor $\sim 2$) by ionizing the gas, thereby reducing the positive charge on grains and PAHs and enhancing FUV-induced grain photoelectric heating. FUV and X-ray photons may create a gap in the disk at $\sim 10$ AU under favourable circumstances.  Photoevaporation timescales for M$_* \sim 1{\rm M}_{\odot}$ stars are estimated to be  $\sim 10^6$ years, after the onset of disk irradiation by FUV and X-rays.  Disk lifetimes do not vary much for  stellar masses in the range $0.3-3$M$_{\odot}$. More massive stars ($\gtrsim 7 {\rm M}_{\odot}$)  lose their disks rapidly (in $\sim 10^5$ years) due to their high EUV and FUV fields. Disk lifetimes are shorter for shallow surface density distributions and when the  dust opacity in the disk is reduced by processes such as grain growth or settling. The latter suggests that the photoevaporation process may accelerate as the dust disk evolves. 

\end{abstract}

\section{Introduction}
Circumstellar disks are closely associated with star formation, with both strong observational evidence for their ubiquitous presence around young stars, and theoretical support as a natural consequence of  angular momentum conservation in  rotating, star-forming cores (e.g., Beckwith et al. 1990, Shu et al. 1994).  Disk  lifetimes are short compared to stellar lifetimes, and  dust disks around solar mass stars are estimated to typically last $\lesssim 3$ Myrs  (e.g., Haisch et al. 2001).  Gas disk lifetimes are not as well constrained observationally ($\lesssim 10^7$ yrs, Zuckerman et al. 1995, Pascucci et al. 2006) but are presumably similar to dust disk lifetimes.  These timescales set an upper limit on the time available for the assembly of planetary systems. Furthermore, dust emission studies that are commonly used to probe disk evolution indicate that disks transition from an optically thick stage  characterised by strong accretion and high ratios of total infrared to bolometric luminosities ($L_{IR}/L_{bol} \sim 10^{-1}$, classical T Tauri star, CTTS) to  a weakly accreting phase with optically thin dust  ($L_{IR}/L_{bol} \lesssim 10^{-4}$, weak-line T Tauri star, WTTS) (e.g., see recent review on dust disk evolution by Watson et al. 2007).  Based on the  statistics of these two  classes of objects, the transition epoch is estimated to be very brief, $\sim 0.1 $Myr, and may involve optically thin inner disks with optically thick outer regions (Hillenbrand 2008, Cieza 2008).  Optically thick gas and dust  disks thus originate during  the star formation process,  undergo significant evolution over a few Myrs  to perhaps also form planets, and are rapidly destroyed during the planet formation process.

Various mechanisms have been proposed to date in order to explain the nature of disk dispersal (see Hollenbach et al. 2000 and Dullemond et al. 2007 for recent reviews).  Disk matter builds planetary systems. However, it has long been recognized from considerations of the total gas and solid content of the solar system and by comparing initial disk masses ($\sim 0.1 {\rm M}_{*}$) to the observed inventory of planet mass ($\sim 10^{-3} {\rm M}_*$) around extrasolar systems that much of the  mass is, in fact, physically removed from the disk either to spiral into the central star or to disperse back into the interstellar medium (e.g., Hayashi et al. 1985).  Viscous accretion is known to occur in the disk and allows transport of  mass onto the central star, but is insufficient as a dispersal mechanism in itself. Viscosity also spreads the disk, as a small amount of matter carries angular momentum outward, and viscous disks do not completely disappear at $\sim$ 10 Myr, as observed (gas masses $\lesssim 10^{-2}$ M$_J$, e.g., Zuckerman et al. 1995, Hollenbach et al. 2005, Pascucci et al. 2006; and dust masses $\lesssim 10^{-4}$ M$_J$, e.g., Beckwith et al. 1990, Haisch et al. 2001). The slow decline in mass of a viscously evolving disk ($M_d \propto t^{-1}$, e.g. Hartmann et al. 1998) is moreover inconsistent with the observationally inferred rapid transition from the CTTS to WTTS phases. One of the most promising mechanisms heretofore proposed for destroying disks has been photoevaporation, where  the  high energy radiation from a young star heats  gas at the disk surface and results in escaping mass flows  from the outer regions of disks.  In this scenario, viscous evolution causes the inner disk to accrete onto the central star and to spread,  while photoevaporation removes the outer disk mass reservoir.

A number of researchers have worked on the problem of disk photoevaporation in various contexts. Hollenbach et al. (1994, hereafter HJLS94),  Yorke \& Welz (1996), and Richling \& Yorke (1997)  examined the photoevaporation caused by the extreme ultraviolet (EUV; $h\nu > 13.6$ eV) photons from the central star. Johnstone et al. (1998), St\"{o}rzer \& Hollenbach (1999) and Richling \& Yorke (1998, 2000) modeled photoevaporation of disks around low mass stars caused by the EUV and the far ultraviolet (FUV;  $ 6 < h \nu < 13.6 $eV) fluxes from nearby massive stars.  Alexander et al. (2004) considered X-rays from the central star and found that they were unlikely to lead to significant photoevaporation rates (also see Ercolano et al. 2008).  Clarke et al. (2001) were the first to combine EUV-induced photoevaporation from a central solar mass star with turbulent viscosity to follow disk evolution. They found that EUV photoevaporation  produces a gap at a characteristic radius $\sim 3-10$ AU once the viscous accretion through the disk has declined to a rate $\dot{\rm M}_{acc} \lesssim 10^{-9}$ M$_{\odot}$ yr$^{-1}$,  below the photevaporation rate from that region for typical EUV fields ($\sim 10^{41-42} {\rm s}^{-1}$). The timescale for producing the gap was estimated as $\sim$  a few Myrs for sufficiently high EUV fields and small initial disk masses. At this point the inner disk rapidly accretes onto the central star on a viscous timescale ($\sim 0.1$ Myr), followed by the  rapid photoevaporation of the outer disk by the unattenuated EUV flux from the central star (Alexander et al. 2005, 2006 a,b).   Matsuyama et al. (2003), Font et al. (2004) and Ruden (2004) have also modeled disk dispersal by a  combination of photoevaporation and viscous spreading and accretion.

Photoevaporation due to the irradiation of a disk by   FUV photons from a central low mass star has not been studied earlier.  FUV and X-ray luminosities of young low mass stars are significant and better determined than their EUV luminosities, which cannot be directly observed. In young low mass stars, FUV, EUV and X-ray fluxes are far in excess of older main sequence low mass stars like the Sun. This is due to both enhanced magnetic activity on stellar surfaces, which leads to much higher chromospheric activity and production of energetic photons, and to the presence of substantial accretion (in accretion columns, Gullbring et al. 1998, Calvet \& Gullbring 1998) onto the central star from the disk, which leads to accretion shocks producing energetic photons.

Although FUV, EUV, and X-rays are produced near the stellar surface, they may be absorbed before reaching the disk surface. Alexander et al. (2005) have shown that the EUV produced in the accretion shock is not likely to penetrate the accretion column.  Disk winds  accompany accretion with a wind mass loss rate that scales as  $\sim \times 0.1$ the accretion rate. In order for the chromospheric EUV  to penetrate the disk wind,  the accretion rate needs to be less than $\sim 10^{-8}$ M$_{\odot}$ yr$^{-1}$ (see Hollenbach \& Gorti 2008). FUV and X-ray photons require much larger column densities than the EUV to be absorbed and therefore begin to penetrate the protostellar wind and  photoevaporate the disk once accretion rates are $\lesssim 10^{-6}$ M$_{\odot}$ yr$^{-1}$ (Hollenbach \& Gorti 2008),  long before the EUV photons from the stellar chromosphere can impact the disk surface.

In this paper, we include the effects of FUV and X-ray radiation from the central star on disk photoevaporation.
This is a preliminary study  and our main goal is to assess the importance of FUV radiation on disk  evolution and lifetimes. We therefore make many necessary simplifications, notably the use of steady-state, static models of the chemical abundances, gas and dust temperatures, and density structure followed by a simple ``streamline'' analysis of the photoevaporative flow. We investigate the relative importance of optical, FUV, EUV and X-ray radiation from the central star in inducing photoevaporation by using our recently developed disk models (Gorti \& Hollenbach 2008, hereafter GH08).  We present models of disks  around central star masses of $0.3-7$ M$_{\odot}$ that span the range of current observational disk detections (e.g., Luhman et al. 2005, Fuente et al. 2006, Manoj et al. 2007).  We consider disks in their initial stages, with moderate grain growth and power law surface density profiles.  We  calculate photoevaporative mass loss rates from the disk as a function of  relevant model parameters  such as the dust opacity per H nucleus, the FUV luminosity, the X-ray luminosity and the power law exponent of the surface density distribution. Although our results are for a snapshot in time, we  qualitatively estimate the effects of viscosity and discuss disk evolution and dispersal.  The  paper is organized as follows.  In \S 2 we present a brief overview of disk evolution prior to the onset of photoevaporation and describe our disk models and photoevaporation calculations.   In \S 3 we discuss our results and  their implications for disk lifetimes, and describe disk evolution as a function of various input parameters.  We summarize and conclude in \S4.

\section{Model Description}
\subsection{Onset of Photoevaporation}
We consider the stage in the star formation process when disk illumination by FUV photons and X-rays is likely to commence.  During the gravitational collapse of the protostellar core and the formation of a central star and circumstellar disk, accretion of mass onto the star proceeds mainly through the disk.  In this earliest phase of stellar evolution,  average accretion rates are very high ($\gtrsim 10^{-6} {\rm \ M}_{\odot}/$yr) and there is complete obscuration of the UV and X-rays from the central star by infall onto the inner disk and the star.  During this phase, disk accretion and transport of angular momentum may be accomplished by global gravitational phenomena such as spiral density waves, as the disk hovers on the brink of gravitational instability.  It is likely that accretion rates through the disk and onto the star are comparable with infall rates onto the disk.  As infall onto the disk ceases, and the disk continues to accrete onto the star, the ratio of disk mass/stellar mass decreases and the self-gravity of the disk becomes less important. Turbulent viscosity may then provide the means of angular momentum transport and the disk mass declines, the disk spreads and the accretion rate on to the star steadily drops (Najita et al. 2007, Bouvier et al. 2007, Watson et al. 2007). Disk illumination by FUV and X-rays commences once they can penetrate the protostellar wind originating in the inner disk. 

Accretion onto the star and stellar magnetic activity provide sources of energetic photons.  There is a significant output of energetic photons from the star due to the accretion hotspots on its surface (FUV, and perhaps EUV and soft X-rays, e.g. Calvet \& Gullbring 1998, Robrade \& Schmitt 2006, G\"{u}del et al. 2007) and due to chromospheric activity (bulk of X-ray emission, e.g, Preibisch et al. 2005 and likely EUV, Alexander et al. 2005).  
  
  Accretion  is also accompanied by  protostellar winds originating from the inner disk that are proportional to the rate of accretion (${\dot{\rm M}}_{\rm w} \sim 0.05-0.1 \dot{{\rm M}}_{\rm acc}$, White \& Hillenbrand 2004). Considerable amounts of  high energy radiation from the star are initially absorbed at the base of the accretion column and in the wind (e.g, Muzerolle et al. 2001).  As accretion diminishes, the opacity of the stellar outflow and the wind from the inner disk is low enough that the outer regions of the disk are irradiated by stellar optical, FUV and X-ray photons. 
 Accretion shock-generated FUV and soft X-ray photons are observationally inferred to at least partially penetrate the accretion column (e.g., Muzerolle et al. 2003, Robrade \& Schmitt 2006), and have been measured around many young stars (e.g., IUE observations for FUV, Valenti et al. 2003). 
  The hydrogen nucleus column density of gas necessary for absorption of X-rays (${\rm N}_{\rm H} \sim 10^{22} {\rm cm}^{-2}$) and FUV (${\rm N}_{\rm H} \gtrsim 10^{22} {\rm cm}^{-2}$, dependent on dust opacity in the protostellar wind) is larger than for EUV absorption (${\rm N}_{\rm H} \sim 10^{20} {\rm cm}^{-2}$). X-rays and FUV photons therefore penetrate the wind and begin to irradiate the disk at earlier epochs than EUV photons. Hollenbach \& Gorti (2008) show that the FUV and X-ray photons begin to penetrate the wind once the accretion rate drops below $\sim10^{-6}$ M$_{\odot}$ yr$^{-1} (\gtrsim
  0.1-1$ Myr), while EUV photons penetrate the wind only when  the accretion rate drops below $ \sim 10^{-8} {\rm \  M}_{\odot}/$yr ($\gtrsim 1-3$Myr).

\subsection{Model Assumptions and Input Parameters}
The main input parameters to the model are the stellar mass, radiation field, the gas phase abundance of elements, the surface density distribution of gas and dust, and the dust properties in the disk.  In this paper,  we calculate the disk structure and photoevaporative mass loss rates at a characteristic time ($\sim 1$ Myr) during the planet-forming disk evolutionary epoch. From the mass loss rates, we estimate disk dispersal times.  We therefore use input parameters such as stellar radiation field and dust properties that are representative of this characteristic time. We list the assumed properties for a fiducial $1{\rm M}_{\odot}$ star case in Table~\ref{fidpar} and for different stellar masses in Table~\ref{starpar}. 

 \paragraph{Stellar Parameters} We assume stellar parameters such as radius, effective temperature and bolometric luminosity as appropriate for a pre-main-sequence star of age $\sim 1$ Myr and use the evolutionary tracks of Siess et al. (2000).  We consider a range of stellar masses from $0.3-7$ M$_{\odot}$.  We do not consider lower mass central objects, although infrared excesses have been noted around objects with masses as low as  $0.1{\rm M}_{\odot}$.  It is rather uncertain as to whether  stars more massive than $\sim 7{\rm M}_{\odot}$ harbor disks around them (e.g., Fuente et al. 2007, Manoj et al. 2007).  In fact, we will show that massive stars ($\gtrsim   7{\rm M}_{\odot}$)  photoevaporate their disks on very short timescales ($\sim 10^5$ years).
The X-ray luminosities of stars are fairly well determined and we use observational data to guide our choice of X-ray luminosity as a function of mass ($L_X \sim 2.3 \times 10^{30} {\rm  (M_*/M_{\odot})^{1.44} erg \ s}^{-1} $ for ${\rm M_* \lesssim 3 M_{\odot}}$ and $L_X \sim 10^{-6} L_*$ for ${\rm M_* \gtrsim 3 M_{\odot}}$;  Flaccomio et al. 2003, Preibisch et al. 2005).  The EUV luminosities of  stars are very poorly known and at best, indirectly determined (e.g., Bouret \& Catala 1998, Alexander et al. 2005).  Here the EUV luminosity (that is not from the photosphere) is assumed to be chromospheric in origin like the X-rays, and of a similar strength and scaling with mass. FUV luminosities of stars are well-studied observationally (e.g., IUE, Valenti et al. 2003, FUSE, Bergin et al. 2003, Herczeg et al. 2004) and theoretically believed to arise due to  both activity in the chromosphere and accretion hotspots on the surface of the star.  We consider both components  for our FUV spectrum.  We calculate accretion-generated FUV luminosities  from mass-correlated accretion rates (e.g., Gullbring et al. 1998 for accretion generated FUV, Muzerolle et al. 2003 for mass-dependent accretion rates; see GH08 for details of our procedure).  The chromospheric component to the FUV flux has an FUV luminosity given by $\log {\rm L}_{FUV}/{\rm L}_* = -3.3$ (data from Valenti et al. 2003), similar to the scaling for X-ray luminosity.  However, as young stars are known to be variable and as disk accretion rates decline with time, we  treat the FUV  and X-ray luminosities as free parameters and vary them over several orders of magnitude, keeping other parameters fixed.

\paragraph{Disk and Dust properties} We assume the initial disk mass to be proportional to the mass of the star with M$_{disk} \sim 0.03{\rm M_{*}}$ (e.g., Andrews \& Williams 2005). Our disk extends from 0.5 to 200 AU and the surface density distribution is generally assumed to follow a power law, $\Sigma(r) \propto r^{-1}$. In one case we assume a $r^{-2}$ power law surface density distribution to probe the sensitivity of the results to the power law exponent. 

 Dust in the disk is assumed to have marginally evolved at epochs of consideration, and we assume a power-law size distribution for the dust with a  range from 50\AA$-20\mu$m, corresponding to a factor of 10 reduction in opacity compared to interstellar dust,  if the gas/dust mass ratio in the disk is  assumed to be 100.   We also consider further reductions in opacity, which we treat by retaining the same size distribution but increasing the gas/dust mass ratio.  Our modeled dust has a mixture of chemical compositions (GH08).  Therefore, at the disk surface where the dust is heated by stellar photons,  the  temperature of a dust grain is a function of both its size and composition.
 
\paragraph{Disk Thermo-chemical Models} Photoevaporative mass loss rates depend sensitively on the density and temperature of the gas in the flow, and in order to calculate X-ray and FUV-initiated photoevaporation rates,  detailed thermo-chemical disk models are required.  EUV photoevaporative flows, on the other hand, are isothermal due to thermal balance processes in ionized gas which set the gas temperature to be nearly constant at $\sim 10^4$K.   Heating and cooling of the predominantly neutral gas below the  ionized surface layer is complex, with dust collisions, X-rays, FUV, photo-reactions and chemistry all being important in determining the density and temperature structure of the gas.  We use our recently developed thermo-chemical, steady-state, disk models  to determine the gas density and temperature structure in the disk (GH08). 

We briefly summarize our disk model here and refer to GH08 for more details.  Our disk models solve for chemistry, thermal balance and impose vertical hydrostatic equilibrium (all self-consistently) to separately calculate the density and gas and dust temperatures as a function of spatial location in the disk. We consider heating of the gas due to X-rays, grain photoelectric heating by PAHs and small grains, cosmic rays, exothermic chemical reactions, formation heating of H$_2$, collisional de-excitation of vibrationally excited (by FUV) H$_2$, photoionization of carbon, and collisions with warmer dust grains.  Cooling of gas is by line emission from atoms, ions and molecules, and by collisions with cooler dust grains. Our chemical network is moderate and focused towards including species that are dominant coolants in the disk (enabling an accurate determination of the gas temperature). We consider  84 species (ions, atoms and molecules) of the elements H, He, C, O, Ne, S, Mg, Fe, Si and Ar and $\sim 600$ chemical reactions, including photoreactions and ionizations caused by cosmic rays.  Our models  neglect ice formation on grains{\footnote{In GH08 we show that in the surface layers where photoevaporative flows originate, a combination of thermal and photodesorption prevent substantial accumulation of ice on grains.}} and calculate steady-state chemistry. We treat gas and dust independently, and allow for different spatial distributions, though we generally consider them well-mixed throughout the disk.  Dust radiative transfer for optically thick dust disks is simplified, so as to keep the numerical disk model computations tractable. We use the two-layer model for dust with some modifications (Chiang \& Goldreich 1997, Dullemond, Dominik \& Natta 2001, Rafikov and de Colle 2006).  We include the effects of  background infrared radiation due to dust on the level populations of the gas species (see Hollenbach et al. 1991).   We then calculate the gas temperature, density and chemical structure as a function of spatial location $(r,z)$ throughout the disk. 

\paragraph{Assumption of steady-state evolution} We emphasize that, in this paper, our entire treatment is {\em steady-state}.  We use stellar properties and radiation fields and disk surface densities at a characteristic age (arbitrarily set at $\sim 1$ Myr), although these are expected to evolve with time as the star evolves.  Our choice for the age of the system is motivated by the observational evidence for disk lifetimes of a few Myrs (e.g., Haisch et al. 2001, Silverstone et al. 2006,  Sicilia-Aguilar et al. 2006) and the expectation that it might be a few $10^5$ years before the star is visible to the disk surface because of the heavy accretion of matter from the dense molecular core onto the star and inner disk (Shu et al. 2000).   The characteristic age of $\sim 1 $Myr therefore approximately corresponds to the onset of FUV and X-ray irradiation of the disk, following penetration of the massive protostellar wind arising from the inner disk.   Disk properties also evolve, with increased dust settling and grain growth as time progresses, but  in these static models we {\em fix} our dust properties, such as the opacity in the disk and the gas/dust mass ratio.  The surface density distribution also evolves under the combined effects of photoevaporation and viscous evolution (e.g. Alexander et al. 2006). However, viscous evolution, if dominant and characterized by a constant viscosity parameter $\alpha$, will tend to maintain a surface density distribution that is approximately proportional to $r^{-1}$. Our aim in this initial study is to estimate the importance of FUV-induced photoevaporation in disks and in this preliminary work we ignore all time-dependent processes. Instead, we focus on deriving {\em average} rates of photoevaporation, by using stellar and disk properties at an intermediate epoch in disk evolution. 
We also use an average EUV field through the disk evolution, although EUV photons may not illuminate the disk until later epochs when the mass loss rate has dropped to $ \lesssim 10^{-8} {\rm \  M}_{\odot}/$yr.  In a future paper, we will couple viscous evolution and photoevaporation in a time-dependent calculation and use evolving stellar fluxes (Gorti, Dullemond \& Hollenbach 2008, in preparation).

We refer the reader to  GH08 for a more thorough discussion on all the above  assumptions and choice of input parameters.

\subsection{Photoevaporation Theory}
Photoevaporative flows are driven by thermal pressure gradients in the gas, and therefore the gas temperature and density distribution (obtained from our disk models) determine the mass loss rates.   Although our models are steady-state  and the disk evolves with time,  we  assume the steady state temperatures and densities do not deviate significantly from the time-dependent values of these physical variables at the surface where the flow originates.  Photoevaporative flows from FUV and X-ray heated surfaces are launched subsonically from regions high in the disk which suffer little extinction to the central star, as will be shown in \S 3. Therefore, heating and chemical timescales are short, of order the timescales for UV photodissociation and X-ray ionization ($\sim 10^{2}$ years at 100 AU), 
 compared to dynamical timescales in the flow ($t_{dyn} \sim r/v_{flow} \sim r/(0.1 c_s) \sim 10^{3-4}$ years at 100 AU). Therefore, the assumptions of steady-state are approximately valid and our calculations provide reasonable estimates of the mass loss rates. The hydrodynamics assumes an isothermal flow from the launch point to the sonic radius, and spherical geometry. Note that the photoevaporative flows we envisage from the surfaces of disks are distinct from the usually bipolar protostellar winds present during the early stages of star formation. We do not treat any possible interaction of the protostellar and photoevaporative winds, which is a complex phenomenon beyond the scope of the present work.  However, we expect the interaction to result in shearing layers which may possibly enhance the photoevaporative mass  loss rates from the disk (Matsuyama, Johnstone \& Hollenbach, in preparation).

The rate of photoevaporation (at a given radius $r$)  is dominated by the layer at a height $z_{flow}$ where the combination of density and temperature yield the maximum mass efflux.  From our calculation of the density and temperature structure of the gas due to heating by FUV, EUV, X-rays and other processes, we find the height $z_{flow}$ where this maximum occurs and thereby obtain the photoevaporation rate as a function of disk radius. As shown by HJLS94, a characteristic gravitational radius $r_g$ enters all discussions of photoevaporation; $r_g$ is the radius where the sound speed or the thermal speed of  hydrogen is equal to the escape speed from the rotating disk. For ionized gas, $r_{g,II} \simeq 7 (10^4 {\rm K}/T) (M_*/1{\rm M}_{\odot})$ AU. In predominantly neutral gas heated by FUV and X-rays
\begin{equation}
r_g \simeq 150 \left( {{1000 {\rm K }}\over{{ T}}} \right) \left({ { { M}_*} \over{1 {\rm M}_{\odot}}}\right) {\rm AU}
\label{rg}
\end{equation}
Note that the gas temperature $T$ varies with $z$, so that it is possible at  a fixed $r$ to have certain heights $z$ where $r<r_g$ and some $z$ where $r>r_g$. 
  If $r>r_g$, then the potential mass flux from a given height $z$ is given by
\begin{equation}
\label{sigmadot}
\dot{\Sigma}(r) = {{d\Sigma(r)}\over{dt} }\sim \mu\ n(r,z)\  c_s(r,z)
\end{equation}
where $\mu$ is the mean particle mass per H nucleus, $n$ is the hydrogen nucleus density, and 
$c_s$ is the sound speed (see HJLS94). 
 
 For lower gas temperatures so that $r<r_g$, Adams et al. (2004) showed that  subsonic flows can be launched. In fact, substantial mass fluxes occur for $r>r_{cr}$ where $r_{cr} \sim 0.1-0.2r_g$ (also see Begelman et al. 1993, Liffman 2003, Font et al. 2004).  The flow accelerates through the sonic point (where the sound speed is attained) and  the photoevaporative flow is launched with the mass flux  given by (see Adams et al. 2004)
\begin{equation}
\label{sigmadot2}
{{d\Sigma(r)}\over{dt}} \sim \mu n_s c_s(r,z) \left( {r_s}\over{r}\right)^2 
\end{equation}
where $n_s$ the density at the sonic radius and $r_s$ is the sonic radius. If we make the assumption that the gas is isothermal from the launch point to the sonic radius, and that $r_s \gg r$ so that spherical symmetry is a reasonable approximation, then
\begin{equation}
\label{dsonic}
n_s = n(r,z) \exp \left( -  {{r_g}\over{2r}}(1-{{r}\over{r_s}})^2\right)
\end{equation}
and $r_s$ is given by 
\begin{equation}
\label{rsonic}
r_s = {{r_g}\over{4}}\left(1+\left(1-{{8r}\over{r_g}}\right)^{1/2}\right)
\end{equation}
for $r<r_g/8$.  A Parker wind solution has $r_s=r_g/2$. For $r>r_g/2$, the flow rapidly goes through a sonic point near the base and $r_s=r$. 
 For $r_g/8 < r < r_g/2$ we linearly extrapolate between $r_s = r_g/4$ at $r=r_g/8$ and $r_s=r_g/2$ at $r=r_g/2$.  
Using Eqs.~(\ref{sigmadot}-\ref{rsonic}) and the density and temperature distribution from our disk models,  the  mass loss rate at each radius is determined by the $z$ layer with the highest  $\dot{\Sigma}$, which could be heated by FUV, EUV, X-rays or any of the other heating processes (\S 2.4). 

 Note that the above analysis assumes that the flow is {\em isothermal} from $r$ to $r_s$, whereas a more accurate hydrodynamical analysis may produce a variation in temperature along a flow streamline. We postpone a more complete analysis to future work and at present check the validity of our solutions by following a radial ray from the star through the disk from the flow surface to the sonic radius. In fact, we find from our thermal balance calculations that the temperature is likely to increase marginally along the flow streamline, increasing the photoevaporation rate from that given by Eqs.~(\ref{sigmadot}-\ref{rsonic}).
 
In the case of EUV-driven photoevaporation, because of the isothermal nature of the ionized region, the flow originates just above the ionization front where the density is greatest. At this height $z_{flow}$, the density is given by (see HJLS94) 
\begin{equation}
n_{II} \simeq 0.3 \left( { {3 \phi_{EUV}} \over{4 \pi \alpha_r r_{g,II}^3}}\right)^{1/2} \left( r/r_{g,II} \right)^{-p}
\label{nhii}
\end{equation}
where $\alpha_r =2.53\times 10^{-13} {\rm cm}^{3} {\rm s}^{-1}$ is the case B recombination coefficient of hydrogen, $r_{g,II} = GM_*/c_{s,II}^2$  is the gravitational radius,  $\phi_{EUV}$ s$^{-1}$ is the EUV photon luminosity of the star  and the exponent $p$  is equal to 1.5 for $r<r_{g,II}$ and
2.5 for $r>r_{g,II}$.  The mass loss rate (or equivalently  $r^2 \dot{\Sigma}$) peaks at $\sim$ $r_{g,II}$ and decreases beyond $r_{g,II}$ as the density at the base $n_{II}$ decreases rapidly with $r$ (HJLS94).  
 
The photoevaporation of gas  occurs at the surface of the disk and carries with it small dust particles. In the absence of turbulence, where dust is allowed to freely settle, the criteria for dust removal is that the terminal speed of a settling dust particle must be less than the speed of the downward moving photoevaporative front. Such a calculation shows that even micron-sized dust particles will settle fast enough to avoid flowing outward with the evaporating surface gas. In other words, the dust abundance in the upper surface layers is extremely low. In the other extreme, one can assume that turbulence mixes the dust vertically
so that no settling occurs. In this extreme, the criterion for dust removal is that the ram pressure of the gas moving upward past the dust is sufficient to accelerate the dust to escape speed. This criterion results in dust with sizes $\lesssim $ mm being coupled to the gas and evaporating while larger dust particles or objects remain behind in the disk (Adams et al. 2004). 
Since  larger particles are decoupled to the flow and may remain in the disk, photoevaporation can decrease the ratio of gas surface density to the dust surface density in disks (see Throop \& Bally 2005 and Youdin \& Shu 2002 for a discussion of the possible consequences). 

\section{Results:  Photoevaporation and Disk Lifetimes}
\subsection{Standard Disk Model for a ${\rm 1 M_{\odot}}$ Central Star}
\begin{figure}
\plottwo{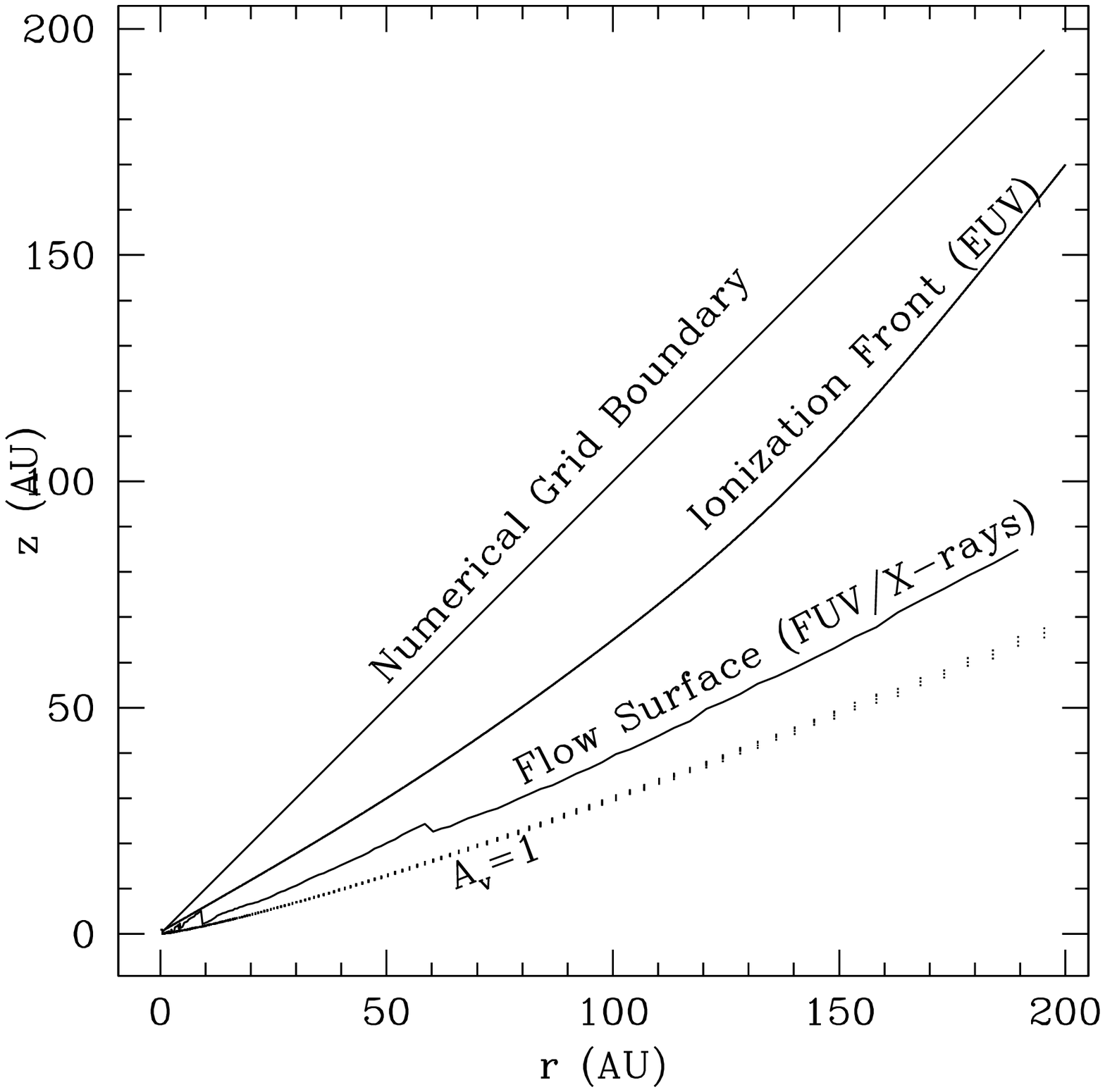}{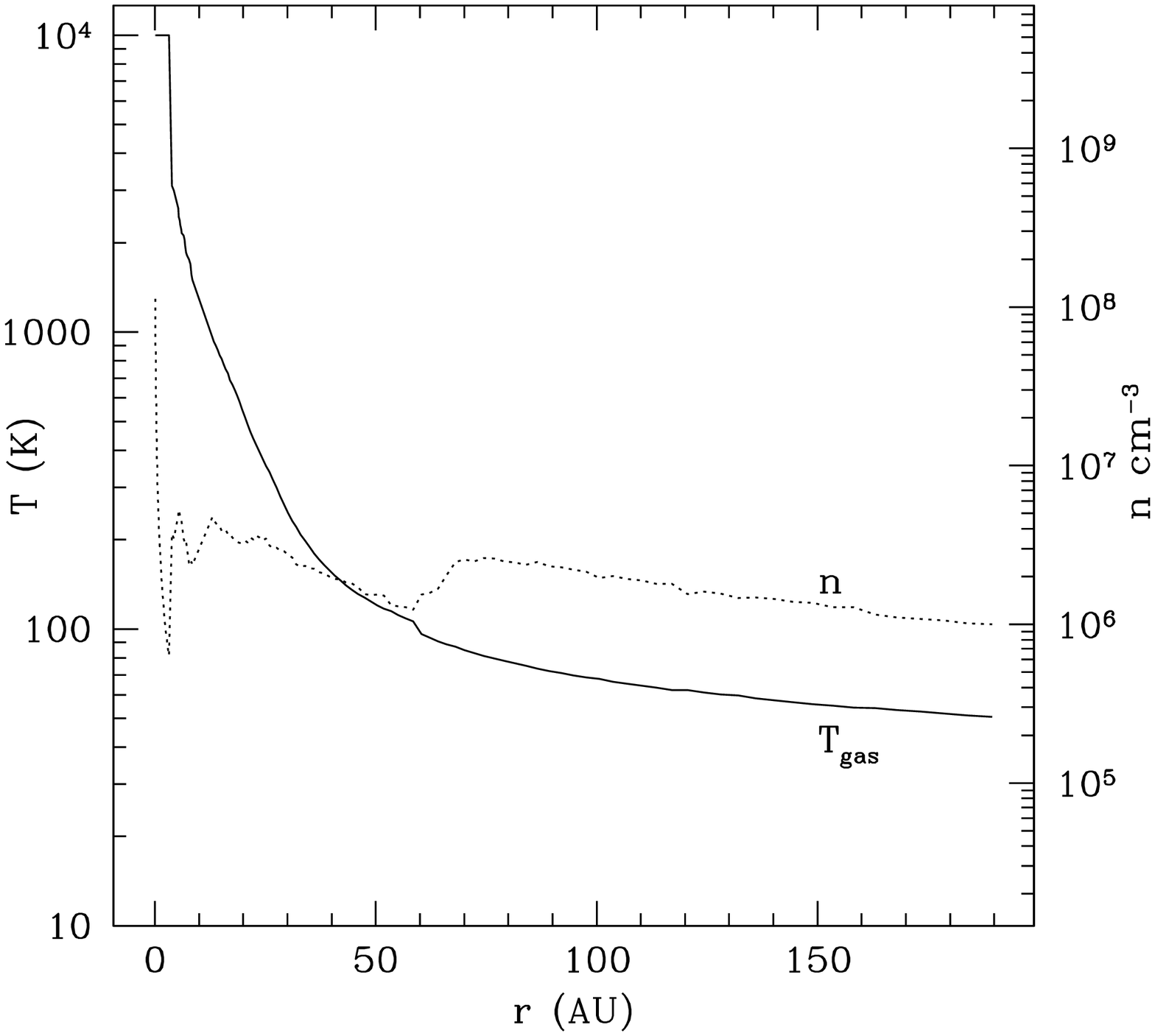}
\caption{Parameters at the flow surface of a photoevaporating disk. In the left panel, the location of the ionization front and the neutral photoevaporative flow surface are marked. The base of the neutral FUV-flow lies slightly above the A$_{\rm V}=1$ layer throughout the disk. The right panel shows the temperature and density at the base of the flow.  For $r<50$ AU, the gas temperature falls steeply with increasing $r$, whereas it is a shallow function of $r$ beyond 50 AU.  The gas density profile shows a sharp kink  at $r\sim3$AU, where the flow shifts from the EUV to the FUV heated region, but is otherwise fairly shallow through the entire disk.}
\label{flowfig}
\end{figure}

We first describe in detail the results for our standard fiducial disk around a ${\rm 1 M_{\odot}}$ star.
(We refer the reader to GH08 for a discussion of the disk structure and chemistry.)  Figure~\ref{flowfig} shows the position of the hot dust surface layer (A$_{\rm V}=1$ to the star at $z=z_{CG}$ in the two-layer dust model, e.g., Chiang \& Goldreich 1997), the ionization front (where the EUV-induced flow would originate) and the FUV/X-ray-induced photoevaporative flow surface for the entire disk. Although we discuss the ionized flow surface and the FUV and X-ray heated neutral flow surface as being distinct for clarity, the true location of the base of the mass flow is at the height where the maximum mass loss (due to EUV, or due to FUV and X-rays) occurs. At the very surface   (A$_{\rm V} \lesssim 10^{-2}$) of the entire disk,  EUV heating (see HJLS94) proceeds via the photoionization of hydrogen  and leads to a nearly isothermal $10^4$ K H II region from the ionization front to the surface.   The gas rapidly becomes neutral and colder below the ionization front. The EUV photoevaporation rate therefore peaks at the base of the  EUV-heated layer, where  the density is highest.  The neutral flow surface ($z_{flow}$), where heating is by FUV/X-rays, is seen to generally trace $z_{CG}$,  but lies  at slightly higher $z$ where A$_{\rm V} \sim 0.01$ for
$r\lesssim 10$ AU,   A$_{\rm V} \sim 0.1-0.2$ for
$r\sim 10-60$ AU, and A$_{\rm V} \sim 0.3$ beyond 60 AU.  This behaviour can be qualitatively understood as follows. The mass loss rate increases with increasing temperature ($T$) and density ($n$), and whereas $T$ increases with $z$ due to greater penetration of FUV and X-ray photons, $n$ decreases with $z$ to maintain vertical hydrostatic equilibrium. The mass loss rate therefore naturally peaks at a height in the gas temperature distribution where both $T$ and $n$ are relatively high. (See GH08 for a discussion of the vertical density and temperature structure in the disk).  This approximately corresponds to the height $z_{CG}$ where the  optical depth due to dust is of order unity and where FUV photons from the star begin to be attenuated, i.e. when A$_{\rm V} \sim 0.1-0.5$.
 
The gas temperature and density at $z_{flow}$, which are the main parameters that disk mass loss rates depend upon, can be approximately estimated by simple considerations.  In order for matter to escape the disk, the gas temperature at the base of the flow has to necessarily satisfy the criterion $r\gtrsim r_{cr}$ with $r_{cr} \sim 0.1-0.2 G M_*/c_s^2$ (see Eq.~\ref{rsonic}, also Liffman 2003, Adams et al. 2004). This implies a criterion for a minimum temperature at $z_{flow}$,  $T(r) \gtrsim  1500 (10 AU/r) 
(M_*/1{\rm M}_{\odot}) $ K.   As the mass loss rate is also a function of gas density which decreases with $z$, the maximum mass loss typically occurs at the lowest height where the minimum flow temperature criterion can be satisfied and the density is still high. Thus, the temperature at $z_{flow}$ is approximately given by the relation $T(r) \gtrsim  1500 (10 AU/r)(M_*/1{\rm M}_{\odot})  $ K.  Simple considerations show that because the flow surface is situated close to ${\rm A}_{\rm V} \sim 0.3$ for most of the disk, the density at $z_{flow}$ lies in the range $\sim 10^6 - 10^7 {\rm cm^{-3}}$.  The hydrogen column density from the star to the ${\rm A}_{\rm V} \sim 0.3 $ surface is $\sim 6\times  10^{21} {\rm cm}^{-2}$ for our standard model. The local density is $\sim 6 \times 10^{21} {\rm cm}^{-2}/r \simeq  4\times 10^6 (100 {\rm AU}/r) {\rm cm}^{-3}$.  Figure~\ref{flowfig} demonstrates the results of a more detailed analysis  using our disk models and  shows the density and temperature  of the gas at the FUV/X-ray flow surface. The gas temperature where the mass loss originates  declines from $\sim 3000$K in the inner ($\lesssim 1$AU) disk to $\sim 1000$K at 10 AU and $\sim80$K at 100 AU in the outer disk.  The density at the base of the FUV/X-ray heated flow does not change appreciably, and remains $\sim  10^6 - 10^7 {\rm cm^{-3}}$ for disk radii 10 AU$<r<$200 AU.

  Heating at the neutral flow surface in the inner disk  ($\sim 3-30$ AU) is mainly due to FUV photons ($\sim$ 70\%), and there is some contribution from chemical processes such as H$_2$ formation heating and X-rays. Cooling is due to [OI]63$\mu$m line emission, H$_2$ vibrational and rotational lines, and dust collisions.  In the outer disk ($r\gtrsim30$AU),  FUV heating still dominates ($\sim$ 80\%) and there is some X-ray heating. Cooling at the flow surface is mainly due to [OI].  Gas temperatures at the flow surface in the outer disk can be below the mean dust temperature, but in general,  the gas is nearly always warmer than the dust at the flow surface so that collisions of gas and dust cool the gas.

\begin{figure}
\plotone{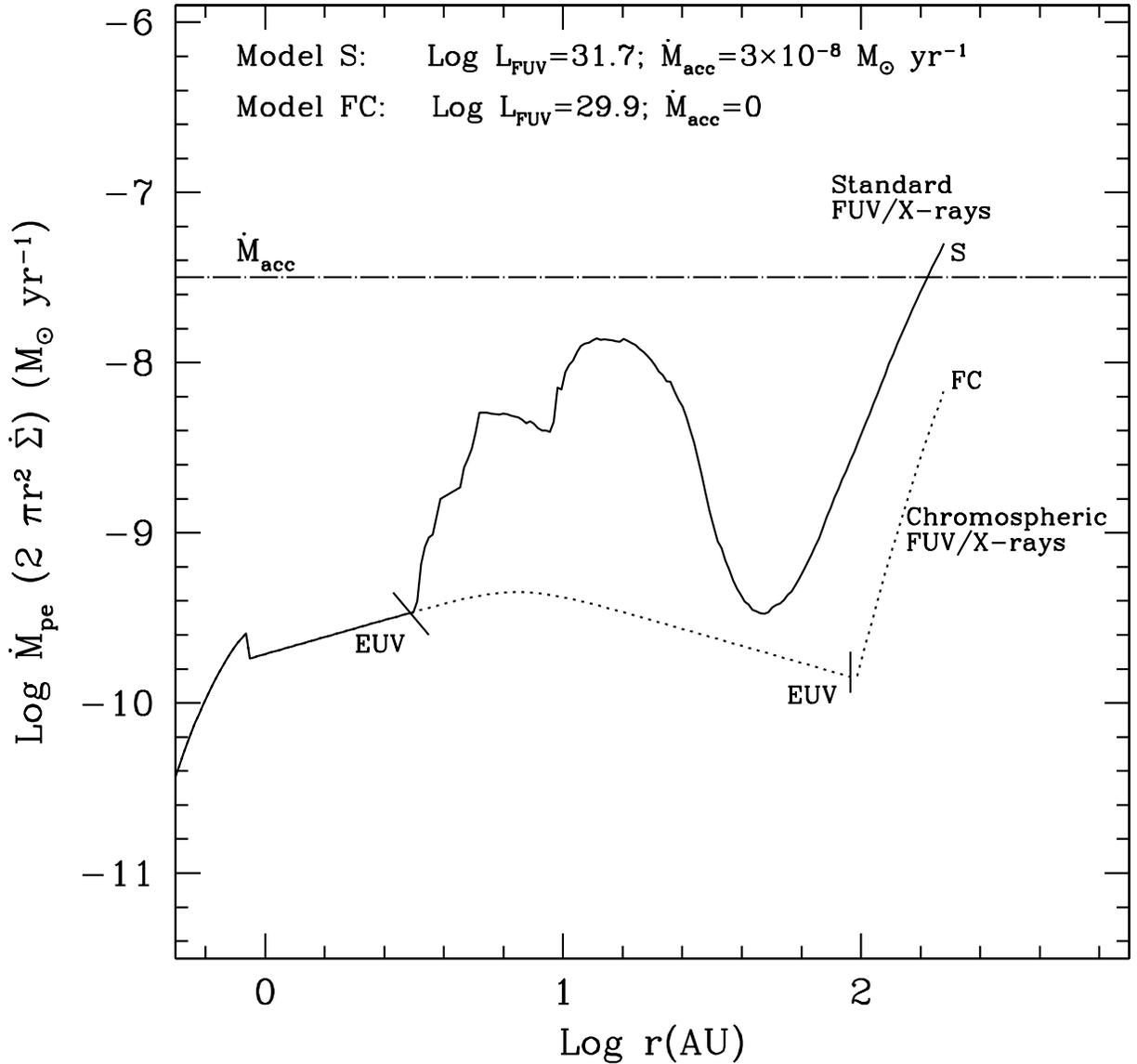}
\caption{The photoevaporation rate expressed as a local mass loss rate ($2 \pi r^2 \dot{\Sigma}$) and shown as a function of radius for our fiducial disk around a ${\rm 1 M_{\odot}}$ star (Model S).  Inside of 3 AU, EUV photoevaporation  dominates, and at $r>3$ AU, FUV/X-ray heating dominates photoevaporation. The mass loss rate peaks at the outer edge of the disk,  and shows a dip at around 40 AU, where the disk is expected to last longest. There is also a peak at $\sim 10$ AU, where a gap may be created at favourable epochs in disk evolution, when $\dot{M}_{pe} > \dot{M}_{acc}$,  the viscous accretion rate (dot-dashed line). The figure also shows $\dot{M}_{pe}$ for another disk model, where accretion has ceased and the FUV is generated by the stellar chromosphere (Model FC). In this case, EUV photoevaporation rates are higher than FUV for $r\lesssim 90$AU, beyond which FUV photoevaporation  dominates.  }
\label{mdotfid}
\end{figure}

Having determined the location of the flow, $z_{flow}$, at every radius as the height with maximum $\dot{\Sigma}(r)$, (Eq.~\ref{sigmadot2}), we next estimate the mass loss rate from the disk.  Figure~\ref{mdotfid} shows the  mass loss rate $\dot{M}_{pe} $ for our standard disk model (Model S, solid line) as a function of disk radius, for  the resulting net (neutral or ionized) flow.   $\dot{M}_{pe}=2 \pi r^2 \dot{\Sigma}(r)$ is approximately the mass loss rate between $r/2$ and $3r/2$, i.e, $\Delta r \sim r$ centered on $r$. 
In the inner disk ($r \lesssim $3 AU),  mass loss rates driven by EUV radiation are higher 
whereas in the outer regions of the disk ($r\gtrsim 3$ AU) FUV/X-ray  photoevaporation is found to dominate.   At $r\sim3$ AU, there is a rapid rise in $\mpe$ as $z_{flow}$ shifts from the ionized region to the neutral, dense FUV and X-ray heated region.    Although mass loss for EUV photoevaporation  peaks at  $\sim r_{g,II}\sim 7 $ AU,  FUV/X-ray induced mass loss rapidly begins to dominate once $r\gtrsim 3$ AU for our standard FUV luminosity which corresponds to an accretion rate of $3\times10^{-8} \mpy$ .  Gas here is heated by FUV and X-rays to temperatures of $\sim 2000$K, and the higher densities in this region result in higher photoevaporative rates than in the $10^4$K, lower density gas in the ionized region. There is another shift in the location of $z_{flow}$ at $\sim 10$ AU, where the flow surface shifts from A$_{\rm V} \sim 0.01$, FUV (80\%) and X-ray (20\%) heated  gas to  denser, predominantly FUV-heated gas at  A$_{\rm V} \sim 0.1$.    At this radius,  Fig.~\ref{mdotfid} shows a corresponding kink in $\mpe$. Beyond $\sim 10$ AU, $z_{flow}$ remains approximately at an A$_{\rm V}$ of 0.3, as discussed above.  However, the surface gas temperature rapidly falls with radius as the FUV/X-ray  flux decreases away from the star and  $\dot{M}_{pe} $ drops sharply at  $\sim  25-40$ AU.  The exponential term in $\dot{\Sigma}$, $e^{-r_g/2r}$ (see Eq.~\ref{sigmadot2}), tends to dominate the mass loss rate in this region. Since $r_g/r \propto (T r)^{-1}$, and because $T$ falls faster than $r^{-1}$ (see Fig.~\ref{flowfig}),  the mass loss rate at $\sim25-40$AU decreases with increasing $r$.  Beyond $\sim 40$AU,  the gas temperature at the flow surface  is  a decreasing shallow function of radius varying as $\sim r^{-0.7}$ and the binding energy of the disk gas begins to decrease,  permitting cooler gas in these regions to escape in neutral flows (decreasing $r_g/r$ factor in Eq.~\ref{sigmadot2}).  Here, although the $e^{-r_g/2r}$ term and $\dot{\Sigma}$ are relatively constant with $r$, $\dot{M}_{pe} \propto r^2 \dot{\Sigma}$ and therefore increases with $r$.  There is a small kink in the location of $z_{flow}$ at $\sim 60$ AU(Fig.~\ref{flowfig}),  where the peak in $\dot{\Sigma}$ shifts from A$_{\rm V} \sim 0.18$ to A$_{\rm V} \sim 0.34$ but $\dot{\Sigma}$ and hence $\mpe$ varies only slightly with $z$ at each radial zone in this region, and  $\mpe$ is relatively smooth with $r$.   

The peak mass loss rates occur at the outer edge of the disk  and are induced by FUV photoevaporation, which therefore sets the disk lifetimes.  Typical surface density distributions in disks are shallow ($\Sigma(r) \propto r^{-a}, a \sim 0.5-1$) such that most of the disk mass is at larger radii (e.g., Andrews \& Williams 2007).   FUV photoevaporation is therefore responsible for the depletion of the disk mass reservoir.  A simple estimate can be made of the  characteristic timescale $t_{pe}$  for FUV-induced photevaporation, by considering the flow parameters in the outer disk.  As shown earlier, $z_{flow} \sim z_{CG}$.  For our assumed dust opacities,  the density at the base of the flow, $n(r,z_{flow})$, can be approximated as   $n(100{\rm AU},z_{flow}) = 4 \times 10^6$ cm$^{-3}$.  At these densities,  thermal balance and chemistry calculations determine the gas temperature as $\sim 80$K. Equation~\ref{sigmadot2} then obtains $\dot{\Sigma} \sim 2 \times 10^{-14}$ gm cm$^{-2}$ s$^{-1}$.  For a 200 AU disk, with mass $0.03$ M$_{\odot}$  and $\Sigma \sim r^{-1}$ the surface density at  100 AU is $\sim$ 1 gm cm$^{-2}$ s$^{-1}$. The characteristic timescale for FUV-induced photoevaporation at 100 AU, where most of the mass resides, is therefore  short,  with $t_{pe} = \Sigma/\dot{\Sigma} \sim  10^6$ years.  
 
\begin{figure}
\plotone{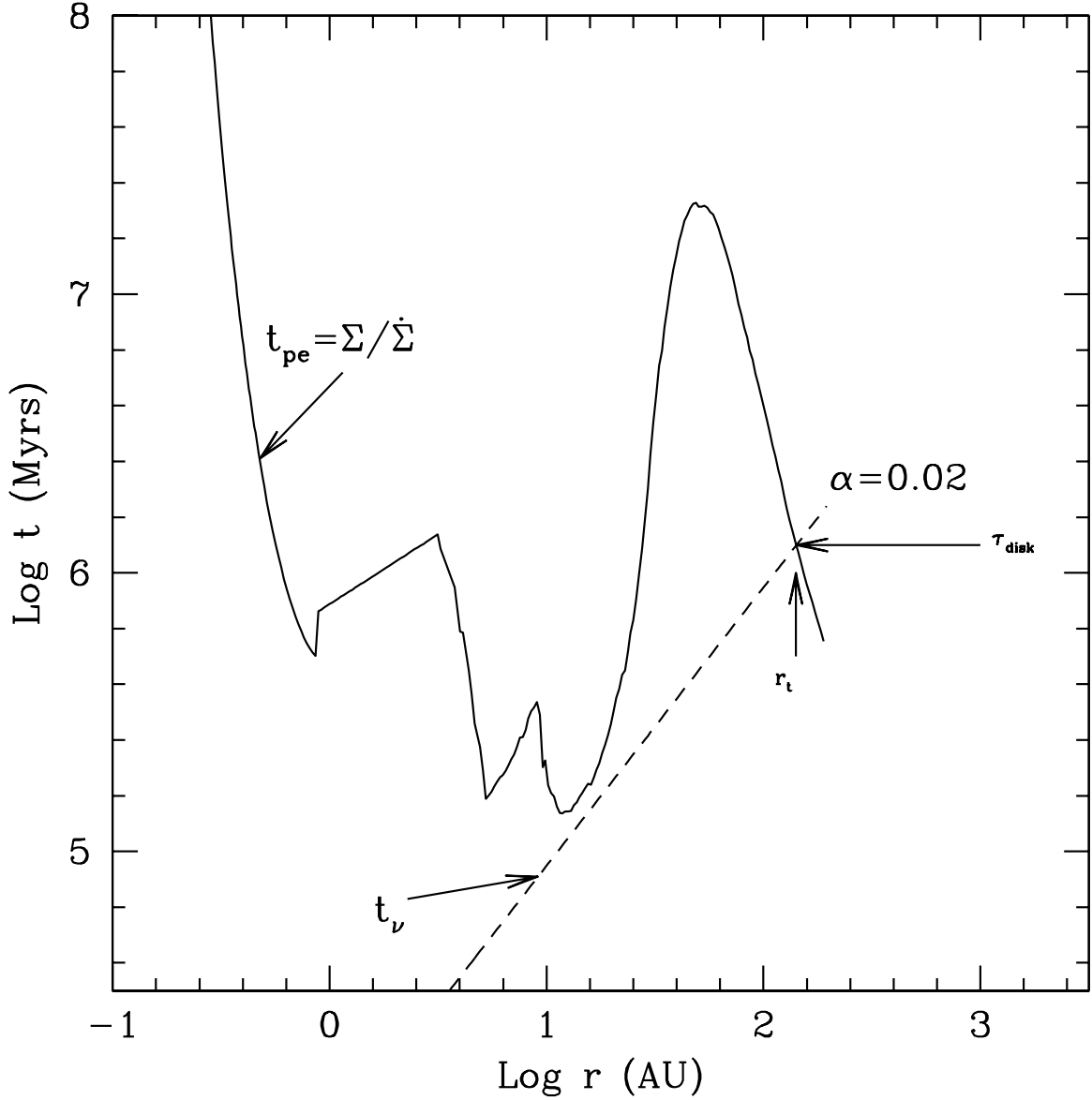}
\caption{The timescale for mass loss due to photoevaporation  ($t_{pe}\sim \Sigma(r)/\dot{\Sigma}(r)$) for our standard disk as a function of radius (solid line). Also shown is the viscous timescale in the disk (dashed line). The disk lifetime $\tau_{disk}$ is estimated as the intersection of $t_{pe}$ and $t_{\nu}$ at some truncation radius $r_t$.  In a time $\tau_{disk}$, the disk mass interior to $r_t$ is drained by viscosity whereas the disk outside of $r_t$  photoevaporates. }
\label{tvis}
\end{figure}

While the disk on the whole begins to lose mass as it is irradiated by FUV and X-ray photons,  the morphological evolution of the disk depends on the relative and local rates of photoevaporation and viscous evolution. Viscosity also plays an important role as it simultaneously spreads the photoevaporating disk. However,  viscous timescales in the outer disk are typically longer than the FUV photoevaporation timescales,  and viscous evolution cannot replenish the photoevaporating  disk at large radii. Therefore, the mass  reservoir here is rapidly drained. In Figure~\ref{tvis}, we show the characteristic photoevaporation timescale, $t_{pe}$ and the viscous timescale $t_{\nu}$ for comparison. 
$t_{pe} (r)$ is calculated as $\Sigma(r)/\dot{\Sigma}(r)$, and viscosity replenishes mass locally on the viscous timescale $t_{\nu} \sim 2r^2/3\nu$, where the viscosity $\nu = \alpha c_s H(r)$, $\alpha$ is the viscosity parameter, $c_s$ is the local sound speed and $H(r)$ is the disk scale height.   We use our disk structure calculations to obtain $H(r)$ and $c_s$.  Our initial assumptions on the surface density distribution with radius and steady mass accretion rate ($\macc = 3 \pi \nu \Sigma$)  in the disk together specify the value of the viscous parameter, $\alpha$ (e.g., Pringle 1981, Hartmann et al. 1998), which for our fiducial disk is calculated as $\sim 0.02$.  Disks may have a range of $\alpha$, depending on the extent of MRI-induced ionization. In \S 3.2 we consider models with the same initial disk mass but varying mass accretion rates (and hence FUV luminosities) which correspond to different values of $\alpha$.  Note that  the value of $\alpha$ does not directly affect $\mpe$ (which depends on the incident FUV luminosity) but does determine the viscous timescale $t_{\nu}$.  We thus calculate $t_{\nu}$ and the two timescales are shown in Fig.~\ref{tvis}.  $t_{pe}$ decreases with $r$ in the outer disk, while $t_{\nu}$ increases with $r$.  A truncation radius $r_t$ can be defined as the radius where these two timescales are equal in the outer disk. For $r>r_t$, $\mpe$ is high, $t_{pe}<t_{\nu}$, and viscous timescales are too long to replace  mass loss by photoevaporation. The disk is therefore rapidly truncated to this radius, and continues to photoevaporate and viscously evolve  while roughly maintaining this size. 
We estimate disk lifetimes by comparing the timescales $t_{pe}$ and $t_{\nu}$  (Figure~\ref{tvis}).  $t_{pe} = t_{\nu}$ at the truncation radius $r_t$.  Mass loss is faster than viscous spreading for  $r>r_t$ in the disk, and viscous accretion removes disk mass at $r<r_t$. The disk loses mass over this characteristic timescale, $t_{pe}=t_{\nu} = \tau_{disk}$, which we define as the disk lifetime. For our fidicual disk, with a constant $\alpha=0.02$,  we obtain a disk lifetime of $10^6$ years after the onset of FUV/X-ray photoevaporation.

Gaps can be created at  a given radius when $t_{pe} (r) < t_{\nu} (r)$, or equivalently, when $\mpe(r) >\macc$.  When $t_{pe} > t_{\nu}$, viscous evolution rapidly transports mass from the outer disk into the local radial annulus, and as long as the outer disk remains massive, photoevaporation does not create gaps in the disk.  For EUV photoevaporation which peaks at  $r\sim r_{g,II}$ (Fig.~\ref{mdotfid}),  when accretion rates in the disk decline to low enough values, a gap forms at $r\sim r_{g,II}$ and thereafter the outer disk evaporates from inside-out (Clarke et al. 2001, Alexander et al. 2006).  FUV/X-ray heated flows may also create gaps at $\sim 3-30$ AU where there is a local peak in   $\dot{M}_{pe}$ (also see Fig.~\ref{tvis}). However, for our fiducial disk,   $\dot{M}_{acc} =3 \times 10^{-8} {\ \rm M}_{\odot} {\rm yr}^{-1}$  and is somewhat greater than the peak value of   $\dot{M}_{pe} $ at $\sim 12$AU, indicating sufficient viscous replenishment of mass to prevent gap formation.  

Creating a gap in the inner disk and maintaining the gap against the smearing effects of viscosity may be a time-dependent phenomenon for FUV-induced photoevaporation. The FUV flux  in our standard model (S) is largely accretion generated,  but as the disk evolves, later lower accretion rate epochs would result in weaker incident FUV fields, and hence also lower $\dot{M}_{pe}$.  Maintaining a gap once it is created therefore depends on how $t_{pe}$ changes as accretion activity decreases with time. If at any instance of time, $t_{pe} < t_{\nu}$  and a gap forms at $r\sim10$AU, then the gas interior to the gap drains onto the star on the viscous timescale here, $\sim 10^5$ years.  However the depletion of the inner disk will result in a lower $\macc$ onto the star, hence lowering the accretion-generated FUV flux and the resulting $\mpe$. In the absence of accretion, a lower limit to $\mpe$ is set by the chromospherically generated FUV flux.  If $\macc$ in the region beyond the gap is higher than this value, then the draining of the inner disk after the first gap creation event would lead to a rapid accretion of the outer disk to fill the inner hole created. The star may perhaps undergo periodic  ``bursts'' of viscous accretion and rapid photoevaporation from the $\sim 10$ AU region at  these epochs, until the accretion rate drops below the lower limit to $\dot{M}_{pe}$ that corresponds to chromospheric FUV and X-ray photoevaporation, or until perhaps EUV photoevaporation becomes significant.   Figure~\ref{mdotfid}  shows the mass loss rate for a model (Model FC;  labels F and C for FUV and chromosphere respectively) where the accretion rate is zero, i.e., the FUV from the star is generated entirely by chromospheric activity. In this case, EUV photoevaporation dominates the inner disk evolution, up to a large disk radius of almost 100 AU, and presumably would maintain (or create) a gap in the disk at $\sim 7$ AU when $\macc \lesssim 5\times 10^{-10} \mpy $. In the outer disk beyond $\sim 100$ AU, heating at the flow surface is by both FUV (50\%) and dust collisions (50\%), and the mass loss rate in the outer disk is still reasonably high, $\sim 10^{-8} \mpy$. We show in \S 3.3 that increasing the X-ray luminosity to $\gtrsim$10 times  its standard value may increase $\mpe$ from FUV/X-rays to a value where it may exceed $\macc$ and hence create a gap.
  
Based on the above results, we propose that disk dispersal around a 1M$_{\odot}$ star proceeds as follows. In the initial stages of evolution, as the accretion rate and hence the  column  density through the protostellar wind from star to disk surface decrease with time, the FUV photons (with absorption columns $\sim 10^{22}$ cm$^{-2}$) begin to irradiate the disk before the wind ceases to be opaque to EUV photons (absorption column $\sim 10^{20}$ cm$^{-2}$).    In the outer regions of the disk, where $\dot{M}_{pe} > \dot{M}_{acc}$, photoevaporation dominates and removes disk material faster than it can be replenished by viscous spreading.  The outer disk regions are eroded rapidly and the disk shrinks.  For plausible surface density distributions, most of the mass of the disk is in the outer regions and much of the disk mass is therefore dispersed via FUV-induced photoevaporation. The disk is expected to rapidly shrink to a truncation radius $r_t$ ($\sim 150$ AU).  Thereafter, viscosity continuously spreads the disk as it loses mass due to FUV photoevaporation, approximately maintaining this outer disk size. Under favourable conditions (for example, high X-ray luminosities), FUV and X-ray photons may create a gap at $3-30$ AU, and cause periodic episodes of inner disk ablation and viscous depletion.  Initially EUV photons have very little effect on the disk evolution, but at later stages penetrate the wind and begin heating the surface.   Finally as the accretion rate declines, a  gap forms in the disk where $\mpe$ due to EUV is a maximum, at $r \sim r_{g,II}$.  At this stage of evolution, most of the disk mass has already been  lost from the disk by FUV/X-ray photoevaporation.  Once the gap forms,   the inner disk rapidly disappears on a viscous timescale of $\lesssim 10^5$ years.  EUV, FUV and X-ray photons now directly impact the newly formed rim  and the outer disk is subsequently  rapidly eroded ($\gtrsim 10^5$ yrs) from {\em inside out}  (as discussed in Alexander et al. 2006).

\subsection{FUV Luminosity} Gas temperatures at the flow surface of the disk are mainly set by the heating due to the  grain photoelectric effect initiated by the incident stellar FUV flux.  Uncertainties and intrinsic variability of the stellar FUV field can therefore affect the photoevaporation rate from a disk.  We investigate the sensitivity of $\dot{M}_{pe}$ and $\tau_{disk}$ to  our assumed FUV field by considering three additional cases. In the earliest stages of evolution, the star actively accretes and can have accretion rates much higher than that assumed in our standard model.  We increase our adopted FUV luminosity by a factor of 10 (corresponding to an accretion rate $\dot{M}_{acc} \sim 3 \times 10^{-7} $M$_{\odot}$ yr$^{-1}$) and calculate the disk structure in this case (Model F10). We also consider a case with the FUV luminosity lowered by a factor of 10 (Model F0.1).  We next consider a disk where accretion has ceased (Model FC).  FUV radiation now stems  only from stellar chromospheric activity and scales with the X-ray luminosity (Wood et al. 2005). Using archival IUE data (Valenti et al. 2003) we find that the median fractional UV excess luminosity for stars with low accretion rates  (we use a procedure similar to Calvet et al. 2004) is given by $\sim$ Log L$_{UV, excess}$/L$_* = -3.3$ and we use this value.  For comparison, we also consider a case where there is no UV or X-rays and the disk is heated only by stellar optical photons (Model O).  All of these models have the same initial disk mass, and hence the derived viscous parameter $\alpha$  for each disk scales directly with the  mass accretion rate, with the fiducial value being $0.02$ (Model S).  

\begin{figure}
\plotone{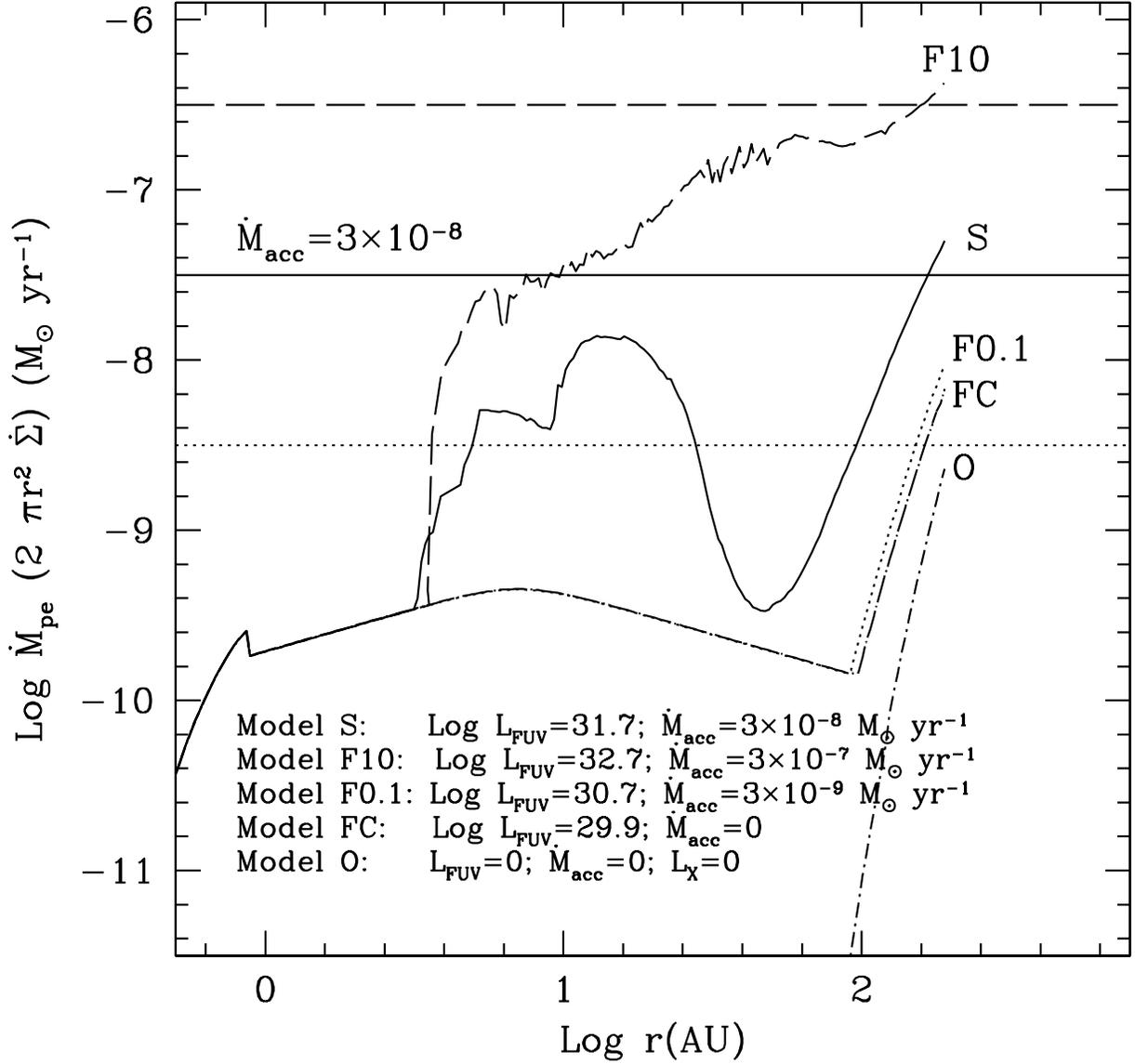}
\caption{ The photoevaporation rates for models with accretion-genererated  FUV  luminosities of  $4\times 10^{30} $(Model F0.1), $4\times 10^{31}$ (Model S),  and
$4\times 10^{32}$ erg s$^{-1}$ (Model F10),  and corresponding mass accretion rates of $3\times 10^{-9}, 3 \times 10^{-8}$ and $3\times 10^{-7}$ M$_{\odot}$/yr respectively. Also shown are a case with no accretion, but only the chromospheric FUV component of $2\times 10^{30}$ erg s$^{-1}$ (Model FC), and a model with only optical photons and no UV or X-rays (Model O). For models F0.1 and FC, EUV photoevaporation dominates for $r\gtrsim 100$ AU.}
\label{mdotuv}
\end{figure}

Photoevaporation rates (and hence derived disk lifetimes)  are quite sensitive to the assumed FUV flux and higher (lower lifetimes) for higher FUV fields.  Figure ~\ref{mdotuv} shows the mass loss rate and the corresponding accretion rate for our different runs.  At earlier stages of disk evolution,  high accretion rates and hence high FUV fluxes incident on the disk will cause rapid mass loss from the outer regions, depleting disk mass significantly. As the disk evolves, both $\macc$ and $\mpe$ decrease, and for the range of FUV luminosities considered here, we find that $\mpe$ decreases faster than $\macc$ for accretion-generated FUV.   From Model F10 to Model S, $\mpe$ decreases by orders of magnitude at $r\sim70$AU, because of the reduced heating.  For both the models F0.1 and FC, EUV dominates for $r\lesssim100$AU, and beyond $\sim 100$ AU FUV and dust collisions heat gas in the flow region. The increased heating causes a factor of $\sim 20$ change in $\mpe$ at $r\sim100$ AU between Models FC and Model O.   At the outer edge of the disk ($r\sim200$ AU), although FUV and dust collisions contribute almost equally to the  heating  for Model FC,  there is not much change in $\mpe$ when compared to the purely optical case (Model O) with only dust heating of gas. The gas temperature at the flow surface is marginally higher for Model FC, but this is compensated by the increased gas density at 
the flow surface in Model O, resulting in only a factor of $\sim 3$ change in  $\mpe$ at 200 AU. 
We therefore find that FUV photoevaporation can cause significant mass loss at early stages of high accretion and associated  high FUV fluxes,  but at later stages when accretion has ceased,  chromospherically generated FUV is not very significant in dispersing the disk. 

Our steady state analysis of the disk around a 1M$_{\odot}$ star with standard X-ray luminosity suggests that FUV/X-ray  photoevaporation may not create a gap in the disk as in EUV photoevaporation.  Although the mass loss rate peaks at $\sim 3-40$ AU in Model S, we find that locally $\dot{M}_{pe}$  is always lower than the corresponding $\dot{M}_{acc}$, indicating that viscous replenishment is always faster and therefore that a gap may never form.  For the same disk, but at different epochs with higher and lower FUV fields, the model results show no evident peak in the FUV-generated mass loss rate.  Thus it appears that in the standard case, a gap almost appears when $\macc\sim 3 \times 10^{-8}\  \mpy$, but not quite. 
 However, note that our models are calculated for fixed epochs and with the standard X-ray luminosity. Time-dependent models that include viscosity  and futher explore the range of X-ray and FUV luminosities are needed to fully assess the manner in which the disk is dispersed by  EUV, FUV and X-ray disk photoevaporation. It may be possible that such models will reveal gap formation at $\sim 10$ AU from the combined effects of FUV/X-rays. In fact, we find below in \S 3.3 that enhanced X-ray luminosities might lead to such a gap.

\subsection{X-rays }
We have up to now treated the combined effects of FUV and X-ray photons in heating the disk gas and causing neutral photoevaporative flows. We next examine in some detail the effect of X-rays on disk heating and photoevaporation.  We consider cases where we vary the stellar X-ray luminosity by  factors of 10 from the fiducial model. In order to isolate the effect of X-rays on disk evolution, we also consider a hypothetical situation where the star produces no EUV  or FUV photons. 

\begin{figure}
\plotone{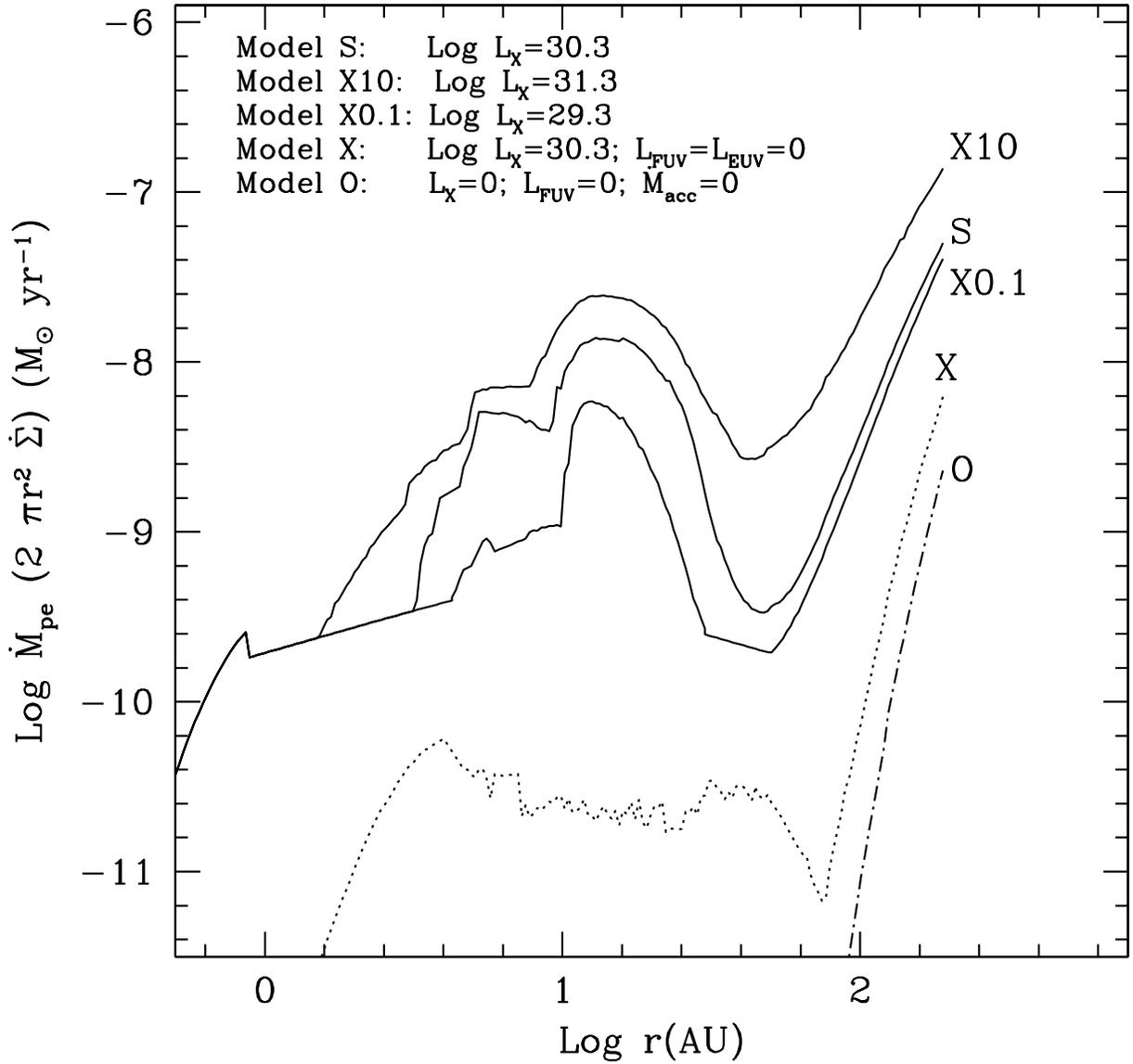}
\caption{ The photoevaporation rate for three different X-ray luminosities, $2\times 10^{31}$ erg s$^{-1}$ (Model X10), $2\times 10^{30}$ erg s$^{-1}$ (fiducial case, Model S) and $2\times 10^{29}$ erg s$^{-1}$ (Model X0.1) as solid lines. All these runs also include FUV ($2\times 10^{31}$ erg s$^{-1}$) and EUV ($2\times 10^{30}$ erg s$^{-1}$)
illumination of the disk. The dotted line shows a hypothetical case (Model X) with no EUV or FUV, but only an X-ray luminosity of $2\times 10^{30}$ erg s$^{-1}$.  The rapid rise in $\mpe$ for model X is not caused by X-rays, but is mainly due to optical photons incident on dust grains which collide with and heat the gas at the disk surface. Disk model O with no FUV or X-rays is also shown for comparison.}
\label{mdotx}
\end{figure}

Figure~\ref{mdotx} shows the results for our three models where $\dot{M}_{pe}$ is shown as a function of radius.  Note that all three runs (S, X10, X0.1) also include EUV and FUV at our fiducial values (Table~\ref{fidpar}). Increasing the X-ray luminosity increases the gas temperature which in turn increases $\dot{M}_{pe}$, as expected.   The effects of X-rays are most pronounced in the inner disk ($r\lesssim 30$AU) where they directly contribute to gas heating. X-rays in combination with FUV can heat the gas to  $\sim 1000-5000$K in the inner disk ($r \lesssim 1-10$ AU), and they can significantly increase the mass loss rates here, by a factor of $\sim 3$ for a factor of 10 change in X-ray luminosity. Note that FUV still dominates the heating here ($\sim 75$\%).  However,  at  small disk radii ($r\lesssim 3$AU), EUV-initiated photoevaporation is higher than that due to FUV and X-rays. The radius at which X-ray heating becomes important decreases as the X-ray luminosity increases, and varies from $1-5$ AU for the range of X-ray luminosities we consider here. X-rays also affect the photoevaporation rates in the outer disk,  by a factor of $\sim 3$ when we increase the X-ray luminosity by a factor of 10 (Model X10) from the standard case (Model S).  Decreasing the X-ray luminosity by a factor of 10 (Model X0.1) does not affect $\mpe$ in the outer disk significantly, as gas heating here is mainly due to FUV photons.  In the inner $r\lesssim 10$ AU disk, however,
decreasing the X-ray luminosity has a more pronounced effect on lowering $\mpe$, which will lower the possibility of gap creation by FUV/X-ray photoevaporation. 

We find that stellar X-rays are largely ineffective in directly dispersing the disk (also see Alexander et al. 2004), although they can significantly affect the heating and chemistry in the gas. In our test run with only X-rays but no FUV or EUV (Model X), X-rays do heat the surface of the gas to $\sim$ few 1000K, but the low densities in these surface regions of the disk result in low mass loss rates from the disk (dashed line in Fig.~\ref{mdotx}).  At $r\sim1-10$AU, where X-rays were earlier found to significantly affect $\mpe$ in our disk models X10, S and X0.1, we find that removing EUV and FUV irradiation has a very dramatic effect. Our model disks S and X have the same X-ray luminosity, but zeroing the UV flux reduces $\mpe$ by a factor of $\sim$ 100. We therefore conclude that X-rays by themselves do not cause significant photoevaporation.   We find that X-rays contribute to the photoevaporation in an indirect manner,  due to the fact that X-ray ionization contributes electrons which thereby increase FUV heating.  FUV heating is mainly by the grain/PAH photoelectric heating which increases as the gas electron abundance increases, since the electrons reduce the positive charge of the grains/PAHs (e.g., Bakes \& Tielens 1994, Weingartner \& Draine 2001). Therefore X-rays act to amplify FUV heating by increasing the electron abundance in the gas.  Even without FUV, Fig.~\ref{mdotx} in case X shows a sudden rapid rise in $\mpe$ at $r\sim 75$ AU, where $z_{flow}$ shifts from low density ($n\sim10^3 {\rm cm}^{-3}$) X-ray heated gas to higher density   ($n\sim10^7 {\rm cm}^{-3}$) gas heated primarily by collisions with dust ($\sim 70$\% of total heating, 30\% by X-rays).  This is clearly seen by comparison with the model disk O (with only optical photon heating) and the addition of X-rays (Model X) increases mass loss only by a factor of $\sim 2$ due to the small increase in temperature. Stellar optical photons, via dust absorption and collisional heating of gas, are primarily responsible for the rise in $\mpe$ at $r\gtrsim75$ AU in Model disk X, and X-rays make only a small contribution to the mass loss here.  The main effect of X-rays on disk photoevaporation is thus indirect, with little direct photoevaporation caused by X-ray heating of gas.

Although ineffective in removing the bulk of the mass in disks, X-rays may have a significant effect on its evolution. In our disk model with increased X-rays (Model X10)  $\mpe \sim \macc$ in the inner  $r\sim10$ AU region of the disk, with $\mpe \sim 2\times 10^{-8} \mpy$. This suggests that a combination of  FUV with high X-rays may be successful in driving a gap in the inner region.  However, as discussed earlier, time-dependent models are needed to validate this tentative result. 

An earlier study  by Alexander et al. (2004) found  X-ray photoevaporation to be unimportant. However, they did not include the effects of FUV heating in their models and only consider direct heating by X-rays  and the resulting photoevaporation.  They find that the upper limit to the rate of change of surface density by X-rays is $\sim 10^{-13}$ g cm$^{-2}$ s$^{-1}$ at 10 AU, comparable to the EUV rate at that radius,  and that most of the X-ray flux is absorbed close to the inner edge of the disk.  A more recent study by Ercolano et al. (2008)  suggests that X-ray photoevaporation may in fact be quite significant, with mass loss rates $\sim 10^{-12}$ g cm$^{-2}$ s$^{-1}$ at 10 AU.  However, they ignore the effects of FUV and EUV, and do not accurately calculate the disk vertical structure. Note that the mass loss rates are very sensitive to the density at the base of the flow (which falls rapidly with $z$ as $e^{-z^2/2H^2}$, $H$ being the disk scale height) and to the gas temperature (exponential dependence of $\dot{\Sigma}$, Eqs.~\ref{sigmadot2}-\ref{dsonic}), necessitating detailed and self-consistent disk structure models.   Our results are more in agreement with the results of Alexander et al. (2004), and we find that X-ray photoevaporation in itself is not very effective in dispersing the disk.   We emphasize that X-rays can nevertheless affect the mass loss rates from the disk via the above indirect effect of raising the degree of ionization in the disk and increasing the efficiency of FUV-induced grain photoelectric heating, and in cases with high X-ray luminosities,  promote the formation of gaps in the inner $r\sim10$ AU disk.

\subsection{Dust Opacity and Surface Density Distribution}
\begin{figure}
\plotone{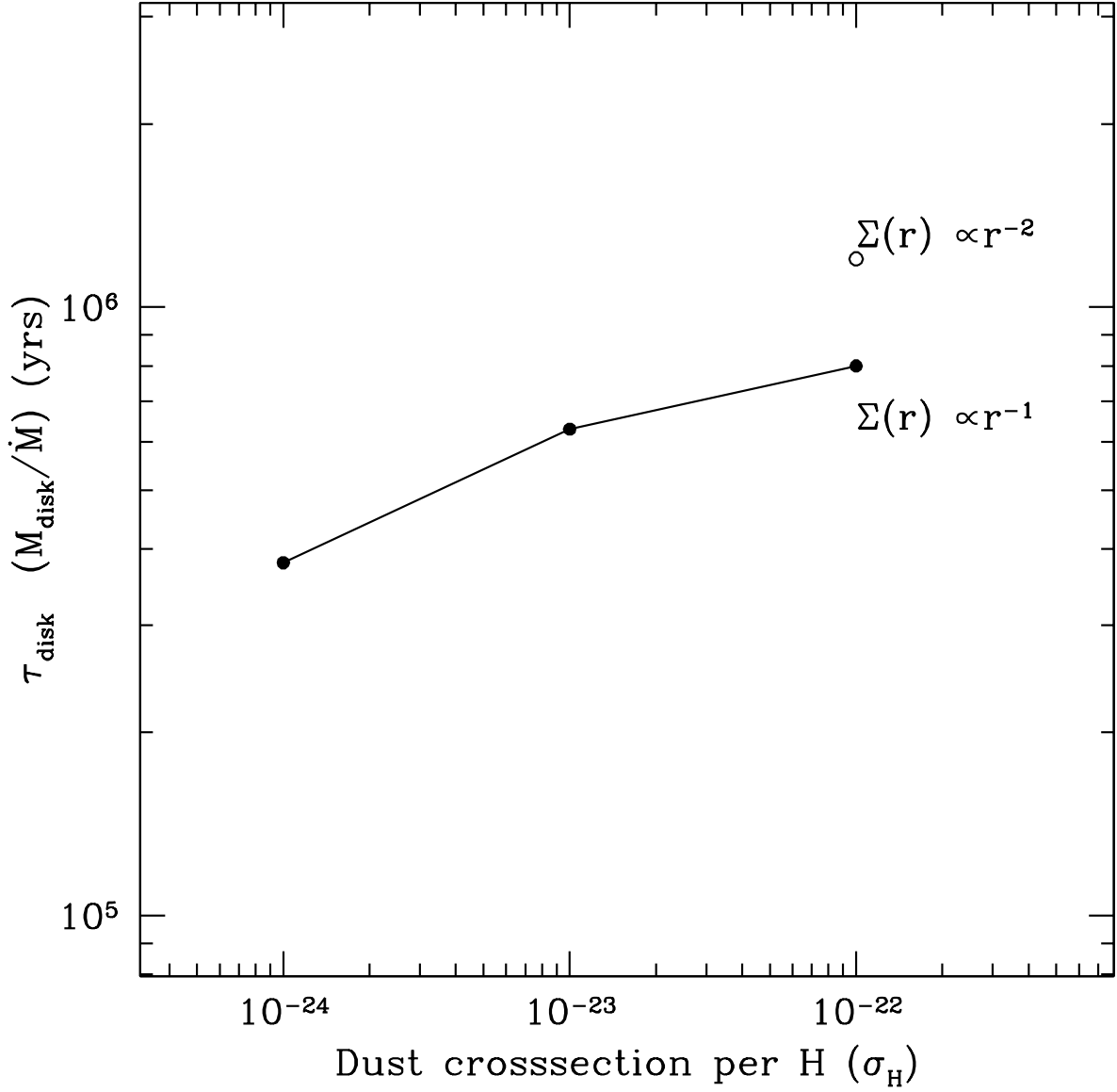}
\caption{Disk lifetime, $\tau_{disk} $  for the 1M$_{\odot}$ case, with different dust opacities in the disk, and the surface density distribution $\Sigma(r) \propto r^{-1}$. Also shown is a case with a dust cross section of $10^{-22}$ cm$^{-2}$  per H nucleus and with a steeper surface density distribution, $\Sigma(r) \propto r^{-2}$.  }
\label{pedust}
\end{figure}

Dust opacity  is an important parameter influencing disk structure, and we investigate how this affects disk photoevaporation rates.  As disks evolve, dust grains are believed to coagulate and grow and perhaps settle (e.g., van Boekel et al. 2005, Natta et al. 2007) reducing the opacity per H nucleus in the disk.  FUV photoevaporation originates close to the ${\rm A_V}=0.3$ layer, and reducing the dust opacity may be expected to increase the photoevaporation rate due to the associated increase in gas density at the ${\rm A_V}=0.3$ layer.  On the other hand, lower dust opacity implies lower FUV heating by the grain photoelectric effect, which would decrease the photoevaporation rate. Dust settling also causes a reduced flaring of the disk (Dullemond \& Dominik 2005),  and hence decreases the amount of high energy radiation intercepted by the disk, perhaps decreasing the photoevaporation rate.    We study these effects by varying the dust cross section per H atom. We keep our dust grain size distribution fixed (from 50 \AA-20$\mu$m), and model a reduction in dust opacity by increasing the gas/dust mass ratio by a factor of 100.  We also decrease the PAH abundance in proportion to the reduction in dust opacity. 
 We find that $\tau_{disk}$ decreases with a reduction in dust opacity (Fig.~\ref{pedust}), suggesting that as the disk evolves and grains settle and grow, the disk might lose its gas even more rapidly.  However, the effect is not large, as seen in  Fig.~\ref{pedust}, with  $\tau_{disk}$  only increasing by a factor of 2, for a factor of 100 increase in dust opacity. Note  that settling timescales depend on disk radius, and are typically longer in the outer disk where FUV photoevaporation dominates.  Our derived disk lifetimes are typically short, $\sim 1$ Myrs, so that dust in the outer disk may not have adequately settled or undergone grain growth to significantly affect disk survival times. 

We also consider a model with a steeper surface density distribution profile with radius ($\Sigma(r) \propto r^{-2}$) compared with the standard disk model ($\Sigma(r) \propto r^{-1}$). We find that  a steeper density distribution increases the disk lifetime, and that this effect is due to  the resulting change in the disk structure.  Steeper surface density distributions cause  less flaring of the optically heated dust disk (e.g., Dullemond  2002),  which results in a smaller angle being subtended by the disk and thereby reduced interception of the FUV and X-ray flux from the star. We note that even steeper profiles may result in the outer disk being partially or completely shadowed by the inner disk,  perhaps limiting the rate at which mass loss can occur.  Photoevaporation of the outer disk may occur until the surface density distribution is steeper than $r^{-2}$ at which self-limiting point the outer disk is shielded by the inner disk, and the photoevaporation rate significantly declines. However, in our present 1+1D dust radiative transfer model,  we cannot investigate this effect and future 2D models are needed. 
 Disks may also be shielded from incident FUV radiation if there are corrugations or ripples in the density distribution caused, for example, by spiral density waves.  We do not consider such effects at present.
 We also note that future time-dependent models are needed to investigate how the surface density distribution evolves as it loses mass due to FUV/X-ray photoevaporation, and how  $\dot{M}_{pe}$ and $\tau_{disk}$ are affected by the changing surface density profile in the disk.

\subsection{Disk lifetimes and stellar mass}

We next investigate the possibility that the disk lifetime is dependent on the mass of the central star.  If disk mass loss rates are determined by FUV photoevaporation due to the central star,  a peak in  disk survival time with mass could possibly arise.    Low mass stars  may have  lower initial disk masses than high mass stars, and their disks are more weakly bound by the stellar gravitational field.  Disk lifetimes may therefore  increase with stellar mass for low to intermediate mass (M$_* \lesssim 3 {\rm M}_{\odot}$) stars.    For massive stars (M$_* \gtrsim 7 $M$_{\odot}$), most of the photospheric stellar emission is in the FUV ( 7 M$_{\odot} \lesssim $M$_* \lesssim 20 $M$_{\odot}$) or in the EUV (M$_* \gtrsim 25 $M$_{\odot}$) and their FUV or EUV luminosities are  high. This may lead to very short disk lifetimes due to photoevaporation.  It may be expected that disk lifetimes gradually increase with $M_*$, peak at intermediate stellar masses ($\sim 3{\rm M}_{\odot}$) and then rapidly drop at higher masses (M$_* \gtrsim 7 $M$_{\odot}$).  Disk lifetimes may have a direct bearing on the likelihood of planet formation in the disk, which may then be a function of stellar mass. 
 We test the hypothesis that disk lifetimes show a peak with $M_*$  with models of disks around stars of different stellar masses, and show below that although disks around high mass stars are short-lived, those around stars from $0.3-3$M$_{\odot}$ have approximately equal lifetimes. 

The FUV luminosity of stars is a highly variable quantity for stars of low mass where the FUV does not arise from the photosphere, and can range over orders of magnitude for a given stellar mass.  In lieu of any definitive observational data characterizing FUV luminosity as a function of stellar spectral type or mass, we use a mass-accretion defined FUV flux as described in GH08, i.e. we use the empirically derived accretion rate vs. mass relation of Muzerolle et al. (2003, 2005) and assume that $\dot{M}_{acc} \propto M_*^2$.  We normalize this to the ``median'' accretion rate for a 1 Myr old solar mass star, $\dot{M}_{acc} = 3 \times 10^{-8}$ M$_{\odot}$/yr.  We then convert $\dot{M}_{acc}$ into an accretion luminosity following the procedures of Gullbring et al. (1998). For massive stars, a significant portion of the FUV flux arises from the photosphere.  It should be noted that due to the uncertain nature of the accretion-generated FUV spectrum, our results are inherently of a qualitative nature.   Table~\ref{starpar} summarizes all the parameters we adopt for the different stellar masses. We additionally include a 30M$_{\odot}$ case where the disk photoevaporation rate is entirely driven by the EUV radiation from the star and use the analytical results of HJLS94 to estimate disk lifetimes.  Recall that we assume the initial disk mass M$_{disk}$ is proportional to the stellar mass, M$_{disk} = 0.03{\rm M}_*$.   This assumption together with the empirical criterion   $\macc \propto M_*^2$,  implies that the viscosity parameter $\alpha$ increases  with stellar mass (see Table 2).

\begin{figure}
\plottwo{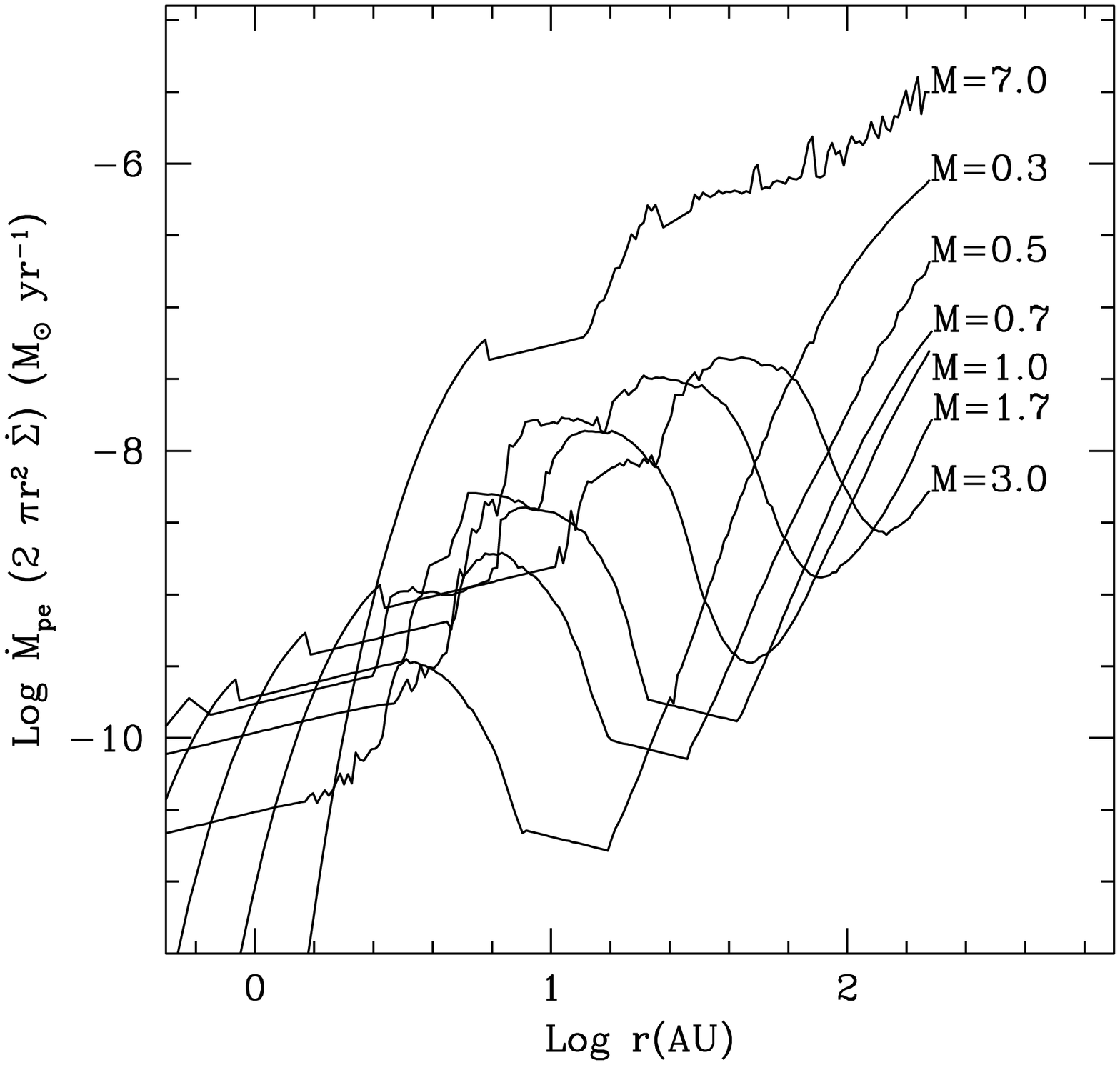}{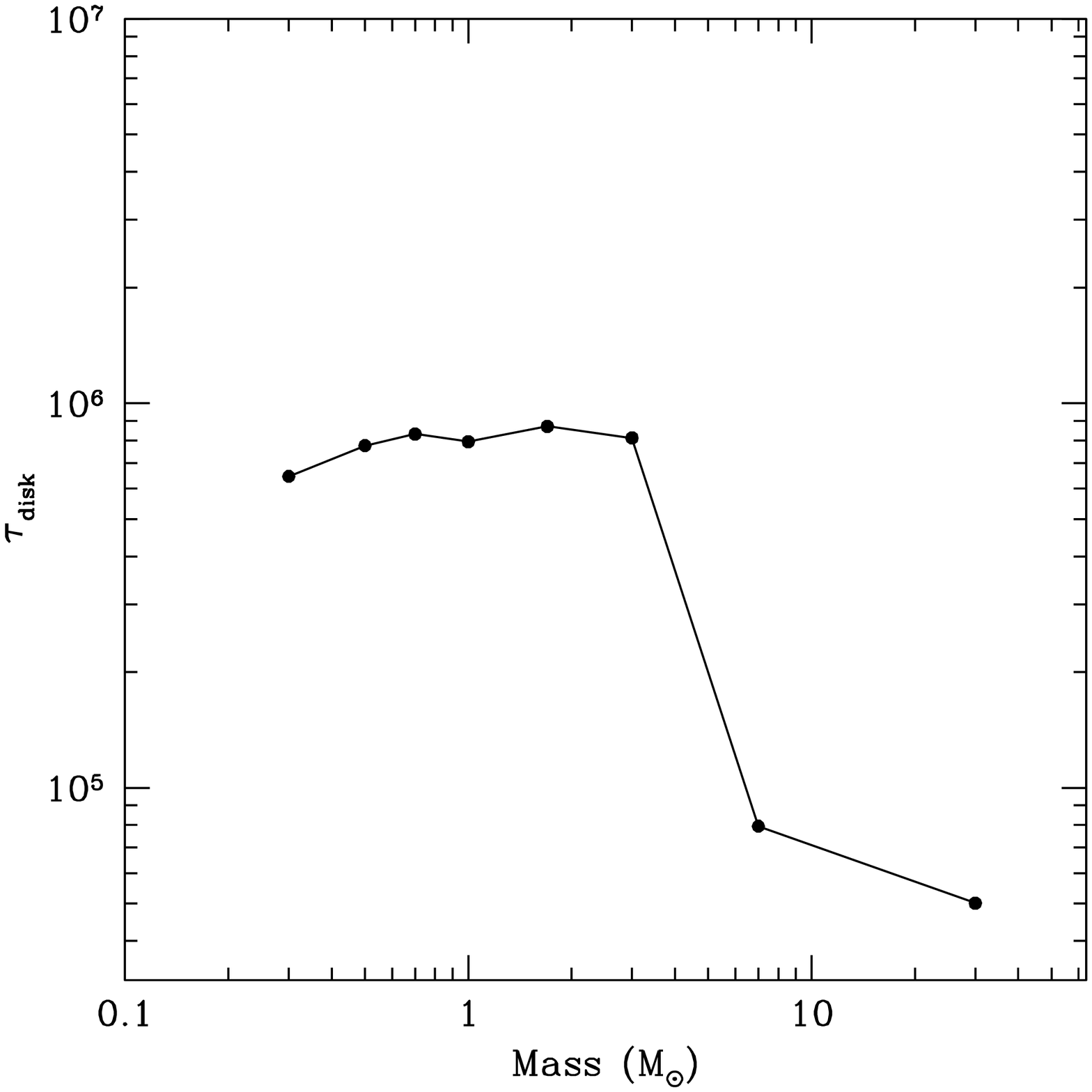}
\caption{Disk mass loss rates $\dot{M}_{pe}$ and derived disk lifetimes are shown for a range of stellar masses.  Disk lifetimes are determined by finding the radius where the photoevaporation timescale $t_{pe}$ is equal to the viscous timescale $t_{\nu}$, and adopting this timescale (see text). Disk lifetimes are nearly independent of stellar mass for $M_* \lesssim 3 {\rm M}_{\odot}$ and sharply decrease  for $M_* \gtrsim 3 {\rm M}_{\odot}$. }
\label{mdotmass}
\end{figure}

Figure~\ref{mdotmass} shows the calculated mass loss rates $\mpe(r)$  for the range of stellar masses considered here.  Evidently, $\dot{M}_{pe}(r)$ does not monotonically increase with stellar mass, even though the FUV luminosity is an increasing function of stellar mass.  There is a peak at the gravitational radius in each case, which shifts to larger radii for more massive stars. The gravitational radius $r_g$  increases with the mass of the star.  This implies that in order to escape the higher gravitational fields,  gas temperature in the flow needs to be higher at a given radius  for higher mass stars.  Gas temperature, however, is only a weak function of the strength of the FUV field.  Therefore, lower mass stars have lower gravitational fields that only loosely binds their lower disk mass, but even with their lower FUV luminosities can heat the gas sufficiently to cause significant photoevaporation. On the other hand, the luminosities of very massive stars are photospheric in origin and significantly higher than accretion-generated FUV luminosities of intermediate and low mass stars, which are a  fraction of the bolometric luminosity.   Strong EUV and FUV fluxes from massive stars can therefore erode their disks rapidly. 
 
 We  find that disk lifetimes  $\tau_{disk}$ do not depend on the mass of the central star for $0.3-3$M$_{\odot}$  stellar masses. We calculate $\tau_{disk}$ as described in \S3.1 for each disk model.  Figure~\ref{mdotmass} also shows the disk lifetime $\tau_{disk}$ as a function of stellar mass for our small sample of disk models. However, $\tau_{disk}$  drops sharply for both massive stars we consider, the 7M$_{\odot}$ star and the 30M$_{\odot}$ star with EUV photoevaporation.  These disks are very short-lived with $\tau_{disk} \sim 10^5$ years.  Our result of a short disk survival time ($\sim10^5$ years) for the most massive stars  is supported by the observational evidence of the  rarity of disks around stars more massive than $\sim$ 7M$_{\odot}$ (e.g., Fuente et al. 2007, Manoj et al. 2007). Massive stars may indeed lose their disks very rapidly due to their strong FUV and EUV fields.  We do not find $\tau_{disk}$ increasing with stellar mass from $0.3-3$M$_{\odot}$ mainly because the FUV  photons that dominate photoevaporation are generated by accretion, characterised by the relation $\macc \propto M_*^2$. 
 Note that the heating at the flow surface in the $0.3$M$_{\odot}$ case (which has very low FUV) is by optical heating of dust grains which then collide with and heat gas. Therefore, even though low mass ($M_*\lesssim 1{\rm M}_{\odot}$) stars start with less massive disks, and these disks are weakly bound, the much lower FUV for the lower mass stars also decreases photoevaporation rates and leads to lifetimes comparable to those for $1-3$M$_{\odot}$ stars.    We conclude with the caveat that this result is very dependent on the various assumptions made in the analysis.  Although our choice of  model input parameters was guided by observational constraints, in many cases the existing data is not  definitive.  A different choice of input parameters, for example, initial disk mass proportional  to $M_*^2$, or   $\macc \propto M_*$,  may result in a peak in disk lifetimes with stellar mass. Our analysis involves assumptions  that affect the value of $\alpha$ in the disk, affecting $t_{\nu}$, and hence the derived disk lifetimes.  Furthermore,  $\alpha$ may vary both as function of disk radius ($r$) as well as disk height ($z$) due to varying ionization levels caused by MRI activity, and we consider a constant $\alpha$ throughout the disk.  Our results are  hence qualitative in nature.  
 
Observational evidence for a peak in disk lifetimes as a function of mass is not very conclusive. There appears to be a peak in disk fraction as a function of stellar type (a surrogate for stellar mass) in some clusters (Lada et al. 2006; Luhman et al. 2008; Strom, private communication), as well as an increased incidence of debris disks with increasing stellar mass (Trilling et al. 2007, Hillenbrand et al. 2008, Cieza et al. 2008).  Carpenter et al. (2006) report disk fractions as a function of stellar mass for the 5 Myr Upper Sco  association and they find that that at least the inner disks (as probed by mid-infrared dust emission) are longer lived for low mass stars ($0.1-1.2$ M$_{\odot}$), and for more massive stars ($1.8-20$ M$_{\odot}$) and are shortest in the intermediate $1-2$ M$_{\odot}$ range.   On the other hand, observations of disks in young clusters such as Tr 37 and $\eta$ Cha (Sicilia-Aguilar et al. 2006, Megeath et al. 2005) indicate that disk fractions increase monotonically with central star mass (K-M spectral type range), and are less likely to be found around lower mass stars.  Recent studies of associations in Orion (Hernandez et al. 2007) indicate a peak at a spectral type of M0, which corresponds to a mass of 0.5M$_{\odot}$.  Damjanov et al. (2007) conclude that the disk frequency in Chameleon I  is not a function of stellar mass, for a spectral type range from K3-M8. An important caveat is that the observational data report dust detections and do not directly trace the gas component or mass of the disk. Conversion of dust observations to disk masses make assumptions about the dust opacity law which is rather poorly known (e.g., Andrews \& Williams 2007) and also assume a gas/dust mass ratio. 

Several studies may help  validate or rule out photoevaporation as the principal agent in disk dispersal.  Further observations of dust disk fractions in clusters may help resolve the seemingly contradictory results obtained so far. New studies that probe gas directly will be helpful.  Better theoretical models are also warranted.  The inclusion of viscosity in  time-dependent disk models (Gorti, Dullemond \& Hollenbach, in preparation) and  perhaps improved  determinations of  UV fluxes to input into these models may also help reconcile the present differences between observational data and model predictions.

\section{Summary and Conclusions}

We  present  results of a study on the effects of  optical, FUV, EUV and X-ray photons from a star illuminating its own circumstellar disk and their role in dispersing the disk gas. Earlier work on photoevaporation focused on the ionizing EUV photons either from a nearby massive star or from the central star in an early energetic phase in stellar evolution. Here we mainly examine the role of FUV and X-ray photons which are produced even by low-mass stars during and after their accretion phase and their effectiveness in driving mass loss from the disk. We include optical, EUV, FUV and X-ray heating from the central star and find that FUV photoevaporation is most efficient in getting rid of bulk of the disk gas, located in the outer regions of the disk. 

We find that FUV photoevaporation induces neutral flows in the disk that are most significant at the outer disk regions ($r\gtrsim100$AU), where most of the disk mass resides.  Typical mass loss rates for a 1M$_{\odot}$ star are $\sim 3\times 10^{-8} $ M$_{\odot} {\rm yr}^{-1}$ at $100-200$ AU, and FUV photons incident on the disk (of initial mass $0.03M_*$) can therefore deplete most of its mass rapidly, on timescales of the order $\sim 10^6$ yrs.  X-rays do not cause significant flows, but can enhance FUV photoevaporation rates by ionizing the gas and increasing the efficacy of FUV grain photoelectric heating of gas.  EUV photoevaporation is effective in the inner ($r\lesssim 3$AU for a 1M$_{\odot}$ star) regions of the disk ($\dot{M}_{pe} \sim 10^{-10} $ M$_{\odot} {\rm yr}^{-1}$) where the stellar gravitational potential is high, and higher temperatures ($10^4$ K) are needed for escaping flows. EUV photoevaporation may also affect disk evolution at later stages by creating gaps in the disk (at $r \sim 7$ AU, 1M$_{\odot}$ case). FUV radiation acting with relatively high X-rays could potentially create gaps at $3-30$ AU at some stages of evolution, but this is not very conclusive from the static model results of this paper.

FUV photoevaporation depletes the disk mass early in its evolution, before EUV photons are able to penetrate the protostellar winds that accompany accretion. We suggest that disk dispersal is initiated by FUV  photons (aided by X-rays) which truncate the disk at the radius where viscous timescales equal photoevaporation timescales (typically at $r\sim 100$ AU) and remove much of the disk mass.  However, time dependent models that follow the decrease in FUV luminosity as the accretion rate drops in time are needed.  As the disk mass drops,   viscous accretion rates decrease to values where EUV photoevaporation can create a gap. This occurs when the EUV-induced mass loss rate at $r_{g,II} \sim 7 (M_*/({\rm 1 M}_{\odot})$ is higher than the accretion rate here.  The inner disk then drains  viscously and the direct illumination of the inner rim (at $\sim 7 $AU) by EUV, FUV and X-rays then results in photoevaporation of the remaining outer disk gas. However, it is the FUV photoevaporation of the outer disk mass reservoir that sets the disk lifetime.  We find that FUV photoevaporation rates (and hence disk lifetimes) are sensitive to the FUV luminosity of the star. If the FUV luminosity is mainly generated by accretion activity, the mass loss rates decline with time as accretion diminishes in the disk.  FUV generated by stellar chromospheric activity is not very significant in dispersing the disk, and produces mass loss rates similar to that by stellar optical photons and dust heating of gas at $\sim 200$ AU.  Disk photoevaporation timescales ($\tau_{disk}$) are calculated at $\sim 1$ Myrs after the penetration of the disk wind by FUV and X-rays (at an age $\sim 1$Myr) and we estimate that disks  may last $\sim2 $ Myrs.

We also find that the disk dust opacity may  influence the photoevaporation rates by affecting
gas density at the flow origin and the degree of flaring.  For a factor of $\sim 100$ reduction in dust opacity, we find that $\tau_{disk}$ decreases by a factor of $\sim 2$.   The surface density distribution of gas in the disk may also affect the photoevaporation rate, as steeper radial distributions decrease the radiation intercepted by the disk and reduce the mass loss rates; changing our surface density from $\Sigma \propto r^{-1}$ to $r^{-2}$ increases the disk lifetime by $\sim$ 50\%.

Disk lifetimes do not appear to depend on the mass of the central star for stellar masses between $0.3$ and 3M$_{\odot}$ stars. Here, we derive lifetimes of $\sim 10^6$ years. High mass stars ($M_* \gtrsim 7$M$_{\odot}$)  lose their disks rapidly in $\sim 10^5$ years because of their very high photospheric FUV and EUV fields. 

Our derived  disk lifetimes  of  $\sim 10^6$ years after the FUV penetration of the protostellar wind for typical solar-mass stars are  consistent with observations, although this study does neglect dynamics and only crudely considers viscosity.  In addition, derived disk lifetimes may be longer if the initial disk mass is greater  than $0.03M_*$.  However,  our qualitative estimates of the ``mean'' disk lifetime by assuming ``mean'' or average stellar properties at a typical age of 1 Myr,  compare very well with observationally inferred disk lifetimes, without requiring very high EUV photon fluxes (Alexander et al. 2006) or special birthplaces near massive OB stars in a cluster environment (e.g., Hollenbach et al. 2000, Adams et al. 2004, Johnstone et al. 2004). 

We offer a concluding important caveat.  The present calculations are simplistic and  only qualitatively include the effects of viscosity.  As the FUV field, accretion rates, and disk properties evolve with time, a proper treatment needs to consider all these effects  in a time-dependent model.  A future goal will be to incorporate the present FUV, EUV and X-ray photoevaporation models into a self-consistent and time-dependent solution that includes viscosity (Gorti, Dullemond \& Hollenbach, in preparation)

\acknowledgements
We would like to thank Richard Alexander, Cathie Clarke, Kees Dullemond,  Steve Strom and  Mark Wolfire for many helpful discussions during the course of this work.   We acknowledge financial support by research grants  from  the NASA Origins of the Solar System Program (SSO04-0043-0032),
Astrophysics Theory Program (ATP04-0054-0083), and the NASA Astrobiology Institute.

\clearpage

\begin{table}
\centering
\caption{Fiducial Disk Model - Input Parameters}
\label{fidpar}
\begin{tabular}{ll}
\\
\tableline
\tableline
Disk mass & 0.03 M$_*$ \\
Surface density & $\Sigma(r) \propto r^{-1}$  \\
Inner disk radius & 0.5 AU \\
Outer disk radius & 200 AU \\
Gas/Dust Mass Ratio & 100 \\
Dust size distribution & $n(a) \propto a^{-3.5} $  \\
Min. grain size $a_{min}$ & $ 50 $\AA   \\
Max. grain size $a_{max}$ & $20 \mu $m  \\
$\sigma_{\rm H}$ & $2 \times 10^{-22} {\rm cm}^2/{\rm H}$ \\
PAH abundance/H  & $8.4 \times 10^{-8}$  \\
\tableline
\end{tabular}
\end{table}

\begin{table}
\centering
\caption{Standard Stellar Input Parameters as a Function of Mass}
\label{starpar}
\begin{tabular}{ccrrcccclc}
\\
\tableline
\tableline
M$_*$ & R$_*$ & T$_{eff}$ & L$_{bol}$ & $\dot{\rm M}_{acc}$ &
 Log L$_{FUV}$ & Log L$_X$ 
& Log $\phi_{\rm EUV}$ & M$_{disk}$ &  $\alpha$\\
(M$_{\odot}$) & (R$_{\odot}$) & (K) & (L$_{\odot}$) & (M$_{\odot}$/yr) & (erg
 s$^{-1}$) & (erg $^{-1}$)  & (s$^{-1}$) & (M$_{\odot}$) \\
\tableline
0.3 & 2.3 & 3360 & 0.55 & 2.7 (-9) & 30.3 & 29.6 & 39.9  & 0.009 & 0.005 \\
0.5 & 2.12 & 3771 & 0.93 & 7.5 (-9) & 30.9 & 29.8 & 40.1 & 0.015 & 0.009\\
0.7 & 2.54 & 4024 & 1.72 & 1.5 (-8) & 31.3 & 30.2 & 40.5  & 0.021& 0.012\\
1.0 & 2.61 & 4278 & 2.34  & 3.0  (-8) & 31.7 & 30.4 & 40.7 &0.03 & 0.02\\
1.7 & 3.30 & 4615 & 5.00 &  8.7 (-8) & 32.3 & 30.7 & 41.0 & 0.051& 0.036\\
3.0 &4.83 & 5004 & 14.85 &  2.7 (-7) &  32.9 &  28.7& 39.0 &0.09 & 0.066\\
7.0 & 3.22& 20527 & 1687 & 1.5 (-6) &  36.5 & 30.8& 44.1 &0.21 & 0.07\\
\tableline
\end{tabular}
\end{table}

\clearpage

\end{document}